\documentclass{article}

\pdfoutput=1

\usepackage{arxiv}

\usepackage[utf8]{inputenc} 
\usepackage[T1]{fontenc}    
\usepackage{hyperref}       
\usepackage{url}            
\usepackage{booktabs}       
\usepackage{amsfonts}       
\usepackage{nicefrac}       
\usepackage{microtype}      
\usepackage{graphicx}
\usepackage{natbib}
\usepackage{doi}

\usepackage{xcolor}
\usepackage{amssymb}
\usepackage{amsthm}
\usepackage{amsmath}
\usepackage[linesnumbered,ruled,vlined]{algorithm2e}
\usepackage{subcaption}

\DeclareMathOperator{\E}{\mathbb{E}}

\usepackage{color}

\DeclareMathOperator*{\argmin}{arg\,min}

\newtheorem{proposition}{Proposition}[section]
\newtheorem{definition}{Definition}[section]
\newtheorem{remark}{Remark}[section]
\newtheorem{lemma}{Lemma}[section]
\newtheorem{cor}{Corollary}[section]

\title{Minimum regularized covariance trace estimator and outlier detection for functional data}

\author{ \href{https://orcid.org/0009-0009-0942-1622}{\includegraphics[scale=0.06]{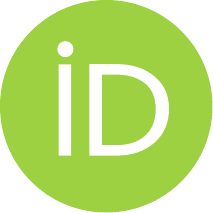}\hspace{1mm}Jeremy Oguamalam} \\
	Institute of Statistics and Mathematical Methods in Economics\\
	TU Wien\\
	Vienna, 1040 \\
 \And
	\href{https://orcid.org/0000-0003-0329-0595}{\includegraphics[scale=0.06]{orcid.pdf}\hspace{1mm}Una Radojičić} \\
	Institute of Statistics and Mathematical Methods in Economics\\
	TU Wien\\
	Vienna, 1040 \\
        \And
	\href{https://orcid.org/0000-0002-8014-4682}{\includegraphics[scale=0.06]{orcid.pdf}\hspace{1mm}Peter Filzmoser} \\
	Institute of Statistics and Mathematical Methods in Economics\\
	TU Wien\\
	Vienna, 1040 \\
}

\renewcommand{\shorttitle}{\textit{arXiv} Template}

\hypersetup{
pdftitle={mrct_manuscript},
pdfsubject={},
pdfauthor={Jeremy Oguamalam, Una Radojičić, Peter Filzmoser},
pdfkeywords={Functional outlier detection, Robust covariance estimator, Regularization, Mahalanobis distance.},
}

\begin{document}
\maketitle

\begin{abstract}
In this paper, we propose the Minimum Regularized Covariance Trace (MRCT) estimator, a novel method for robust covariance estimation and functional outlier detection. The MRCT estimator employs a subset-based approach that prioritizes subsets exhibiting greater centrality based on the generalization of the Mahalanobis distance, resulting in a fast-MCD type algorithm. Notably, the MRCT estimator handles high-dimensional data sets without the need for preprocessing or dimension reduction techniques, due to the internal smoothening whose amount is determined by the regularization parameter $\alpha>0$. The selection of the regularization parameter $\alpha$ is automated. The proposed method adapts seamlessly to sparsely observed data by working directly with the finite matrix of basis coefficients. An extensive simulation study demonstrates the efficacy of the MRCT estimator in terms of robust covariance estimation and automated outlier detection, emphasizing the balance between noise exclusion and signal preservation achieved through appropriate selection of $\alpha$. The method converges fast in practice and performs favorably when compared to other functional outlier detection methods.
\end{abstract}


\section{Introduction}

\label{sec:intro}
The analysis of functional data is becoming increasingly important in many fields, including medicine, biology, finance, and engineering, among others. Functional data are characterized by measurements that are taken over a continuum, such as time or space, and are often modeled as curves or surfaces. One important aspect of functional data analysis is the estimation of the covariance structure, which describes the dependence between different points on the continuum. The covariance structure is essential for various tasks, such as smoothing, regression, and classification, and is often used to identify functional outliers, which are observations that deviate from the expected behavior.

However, the estimation of the covariance structure for functional data is challenging, especially in the presence of outliers or other sources of noise. Traditional covariance estimators, such as e.g. the sample covariance, are sensitive to outliers and may produce unreliable results. Therefore, the development of robust covariance estimators that can handle outliers and other types of noise is an active area of research in functional data analysis.

Several robust covariance estimators have been proposed for functional data in the literature. These estimators are based on different approaches, such as trimming, shrinkage, or rank-based methods, and they aim to produce estimates that are less sensitive to outliers and other sources of noise. They can be combined with outlier detection techniques to identify functional outliers and improve the overall quality of the analysis; see e.g. \cite{Boente2021}, \cite{Locantore1999}, \cite{Cuevas2007}, \cite{gervini2008}, \cite{sawant2012}. In this paper, we propose a novel robust covariance estimator based on a trimmed-sample covariance. The approach is motivated by the multivariate trimmed covariance estimator chosen such that it minimizes the Kullback-Leibler divergence between the underlying data distribution and the estimated distribution. 
The organization of the paper is as follows: Section \ref{sec:motivation} provides motivation for the estimator in the multivariate setting. Section \ref{sec:notation} introduces the notation and presents important properties of the covariance operator in the Hilbert space of square-integrable functions defined on an interval in $\mathbb{R}$. The MRCT estimator is introduced in Section \ref{sec:MRCT}, followed by a discussion of its algorithmic properties and computational aspects in Section \ref{sec:Algorithm}. For handling sparsely observed data, the implementation of the method is presented in Section \ref{sec:sparsedata}. To assess its performance, we conduct an extensive simulation study in Section \ref{sec:Simulations}, comparing the estimator with existing methods across various simulated datasets, thereby demonstrating its effectiveness in handling outliers and improving estimation accuracy. Finally, in Section \ref{sec:Examples}, we apply the MRCT estimator to real-data examples, providing practical illustrations that conclude the paper. Proofs of the technical results are given in Appendix \ref{sec:appendix2}

\section{Motivation}\label{sec:motivation}

In the multivariate setting, given the random sample $\textbf{x}_1,\dots,\textbf{x}_n\in\mathbb{R}^p$ from a distribution with mean $\boldsymbol{\mu}$ and covariance $\boldsymbol{\Sigma}$, one approach to robustify the estimation of the mean and the covariance is to use the weighted sample estimators, which assign different weights, $w_1, \dots,w_n\geq 0$, to different data points based on their relative importance. 
In this approach, outliers or noisy data points ought to be given smaller weights, thus reducing their impact on the estimated mean and covariance. Especially, choosing weights $w_i\in\{0,1\}$, $i=1, \dots,n$, with the constraint $\sum_{i=1}^nw_i=h$, for some fixed $n/2\leq h\leq n$, yields trimmed mean and covariance estimators. For example, one of the most widely used members of this class of estimators is the so-called \textit{minimum covariance determinant} (MCD) estimator \citep{rousseeuw1999}. For a  given fixed number $h$, which is roughly half of the sample size $n$, its objective is to select $h$ (out of $n$) observations for which the calculated sample covariance matrix has the lowest determinant. 
In the parametric family of distributions parameterized by its mean and covariance, we employ a similar strategy, however with the objective to minimize the Kullback-Leibler (KL) divergence between the theoretical (underlying data distribution) and the one parameterized by the estimates of mean and the covariance, using corresponding sample estimates calculated at a  subset $H\subset \{1,\dots,n\}$, $|H|=h$. The KL divergence (also known as relative entropy) between continuous distributions $P$ and $Q$ with densities $f_P$ and $f_Q$, respectively, denoted as $\mathrm{KL}(P\parallel Q):=\E_Q(\log(f_P(\textbf{x})/\log(f_Q(\textbf{x})))$, is a measure of how far the probability distribution $P$ is from a second, reference probability distribution $Q$~\cite{Kullback1951,kullback1959}. The intuitive and straightforward interpretation of the corresponding KL divergence is the expected excess surprise from using $P$ to model the data when the actual distribution is $Q$. As it involves (the estimation) of the density function, direct use of KL divergence is usually mitigated by the use of appropriate approximations; see e.g.~\cite {Blei2017} for more details. 
Assume now the data $
\{\textbf{x}_1, \dots,\textbf{x}_n\}$ originates from the normal $\mathcal{N}(\textbf{0},\boldsymbol{\Sigma})$ distribution, and denote $\bar{\textbf{x}}_H$ and $\hat{\boldsymbol{\Sigma}}_H(\textbf{x})$ to be the mean and the covariance of $\{\textbf{x}_i;\,i\in H\}$. 
Denote further  $\hat{\boldsymbol{\Sigma}}_{H}^{\mathrm{NC}}(\textbf{x})=\frac{1}{h}\sum_{i\in H}\textbf{x}_i\textbf{x}_i'$ to be the "non-centered" covariance of an $h$-subset $H$. The KL divergence between $\mathcal{N}(\textbf{0},\boldsymbol{\Sigma})$ and $\mathcal{N}(\bar{\textbf{x}}_H,\hat{\boldsymbol{\Sigma}}_{H}(\textbf{x}))$ is 
\begin{equation}\label{eq:KL}
\mathrm{KL}(\mathcal{N}(\textbf{0},\boldsymbol{\Sigma})||\, \mathcal{N}(\bar{\textbf{x}}_H,\hat{\boldsymbol{\Sigma}}_{H}(\textbf{x})))=\mathrm{tr}(\boldsymbol{\Sigma}^{-1}\hat{\boldsymbol{\Sigma}}_{H}^\mathrm{NC}(\textbf{x}))+\log(\det({\boldsymbol{\Sigma}}^{-1}\hat{\boldsymbol{\Sigma}}_{H}(\textbf{x})))-p;
\end{equation}
see \cite{Duchi2016} for more details. 
As the second term $\log(\det({\boldsymbol{\Sigma}}^{-1}\hat{\boldsymbol{\Sigma}}_{H}(\textbf{x})))$ in \eqref{eq:KL} is a determinant of a regular matrix on a logarithmic scale, i.e.~sum of the logarithms of the corresponding eigenvalues, we can say that the first term $\mathrm{tr}(\boldsymbol{\Sigma}^{-1}\hat{\boldsymbol{\Sigma}}_{H}^\mathrm{NC}(\textbf{x}))$ dominates \eqref{eq:KL}. 
To support the claim that the minimizer of $\mathrm{tr}(\boldsymbol{\Sigma}^{-1}\hat{\boldsymbol{\Sigma}}_{H}^\mathrm{NC}(\textbf{x}))$ approximately minimizes the KL divergence \eqref{eq:KL}, we can examine the second term $\log(\det({\boldsymbol{\Sigma}}^{-1}\hat{\boldsymbol{\Sigma}}_{H}(\textbf{x})))$ in a different form. First, observe that $\log(\det(\boldsymbol{\Sigma}^{-1/2}\hat{\boldsymbol{\Sigma}}_H(\textbf{x})\boldsymbol{\Sigma}^{-1/2}))=\log(\det(\hat{\boldsymbol{\Sigma}}_H(\textbf{y})))$, where $\hat{\boldsymbol{\Sigma}}_H(\textbf{y})$ represents the covariance of standardized observations ${\textbf{y}_i=\boldsymbol{\Sigma}^{-1/2}\textbf{x}_i:i\in H}$. For $\textbf{y}_i$ we have that $\mathrm{Cov}(\textbf{y})=\textbf{I}_p$, indicating that any strongly consistent estimator $\hat{\boldsymbol{\Sigma}}(\textbf{y})$ of $\mathrm{Cov}(\textbf{y})$ admits the representation $\hat{\boldsymbol{\Sigma}}(\textbf{y})=\textbf{I}_p+\varepsilon_n\textbf{M}_n$, where $\textbf{M}_n$ is a unit-norm $p\times p$ matrix and $\varepsilon_n\to_{a.s.}0$. 
This holds in particular for the consistent estimator $\hat{\boldsymbol{\Sigma}}_{H}(\textbf{y})=\textbf{I}_p+\varepsilon_n\textbf{M}_n$, for some unit-norm $p\times p$ matrix $\textbf{M}_n$. A slight modification of Jacobi's formula implies that the directional derivative of the determinant in the direction of $\textbf{M}_n$ evaluated at the identity  equals the trace of $\textbf{M}_n$, i.e. $\lim_{\varepsilon_n\to 0}\left(\det(\textbf{I}_p + \varepsilon_n \textbf{M}_n) - \det(\textbf{I}_p) \right)/\varepsilon_n = \text{tr}(\textbf{M}_n)$.
The continuity of the determinant along with applying the identity to the first order Taylor expansion of $\det(\hat{\boldsymbol{\Sigma}}_{H}^{\mathrm{NC}}(\textbf{y}))$ around the identity yields that for some $\delta_n\to_{a.s.}0$
\begin{align*}
    \mathrm{det}(\hat{\boldsymbol{\Sigma}}_H(\textbf{y}))=&~ \mathrm{det}(\hat{\boldsymbol{\Sigma}}_H^\mathrm{NC}(\textbf{y}))+\delta_n= \det(\textbf{I}_p) + \varepsilon_n\text{tr}(\textbf{M}_n) + o(\varepsilon_n^2)+ \delta_n \\
 =&~\mathrm{tr}(\boldsymbol{\Sigma}^{-1}\hat{\boldsymbol{\Sigma}}_{H}^\mathrm{NC}(\textbf{x}))+(p-1)+o(\varepsilon_n)+\delta_n.
\end{align*}

This result implies that for sufficiently large $n$, the determinant of the consistent estimator of the covariance of the standardized observations behaves similarly to its trace, thereby reinforcing the claim that minimizing the trace of the non-centered covariance of the standardized observations approximately minimizes \eqref{eq:KL}.
Thus, instead of directly minimizing $\mathrm{KL}(\mathcal{N}(\textbf{0},\boldsymbol{\Sigma}), \mathcal{N}(\bar{\textbf{x}}_H,\hat{\boldsymbol{\Sigma}}_{H}(\textbf{x})))$ over all subsets $|H|=h$, we consider the following optimization problem
\begin{equation}\label{eq:MVT_theorethical}
H_0=\argmin_{H\subset\{1,\dots,n\}:|H|=h}\mathrm{tr}\left(\hat{\boldsymbol{\Sigma}}_{H}^\mathrm{NC}(\textbf{y})\right)=\argmin_{H\subset\{1,\dots,n\}:|H|=h}\sum_{i\in H}\|\textbf{y}_i\|^2,
\end{equation}
where $\hat{\boldsymbol{\Sigma}}_{H}^\mathrm{NC}(\textbf{y})=\frac{1}{h}\sum_{i\in H}\textbf{y}_i\textbf{y}_i'$ is a non-centered covariance of standardized observations $\{\textbf{y}_i=\boldsymbol{\Sigma}^{-1/2}\textbf{x}_i:i\in H\}$. The corresponding estimators $\hat{\boldsymbol{\Sigma}}_{H_0}(\textbf{x})$ and $\bar{\textbf{x}}_{H_0}$ of the covariance and the mean are then referred to as multivariate Minimum Covariance Trace  (MCT)  estimator of the mean and the covariance, respectively.

The optimization problem \eqref{eq:MVT_theorethical} highlights the need for an appropriate standardization of the data when extending this approach to more general spaces. The standardization ensures that the norm of the standardized object reflects a specific dissimilarity measure of the object from its corresponding mean. In the multivariate setting, where $\boldsymbol{\Sigma}$ is a positive definite matrix, the norm of the standardized observations $\textbf{y}_i$ corresponds to the Mahalanobis distance between the original observation $\textbf{x}_i$ and the mean. Thus, we aim to find the $H$-subset, $|H|=h$, 
as a solution to a generalized form of problem \eqref{eq:MVT_theorethical}. However, before delving into the functional setting, it is important to address the limitations of using the trace of raw observations as the objective function.

\begin{remark}\label{rem:rem1}
If the data were not standardized, the method would prioritize observations with the smallest Euclidean distance from the center of the data cloud. This would introduce a significant bias towards selecting spherical data subsets for covariance estimation. Conversely, if the data already originated from a spherical distribution, or at least exhibited an "approximate" spherical behavior, this drawback would be mitigated.
\end{remark}

\section{Notation and preliminaries}\label{sec:notation}

Let $X$ be a random function on $L^2(I)$, i.e.~the Hilbert space of square-integrable functions over the interval $I$. The corresponding covariance function is defined as $v(s,t)=\mathbb{E}((X(s)-\mu(s))(X(t)-\mu(t)))$ and is continuous, symmetric and positive semi-definite. Here, $\mu(t)=\mathbb{E}(X(t))$ with $t \in I$ denotes the mean of $X$. Let further $C:L^2(I)\to L^2(I)$ be the corresponding covariance operator defined by $\displaystyle Cx(t)=\mathbb{E}\left(\langle x,X-\mu\rangle \left(X(t)-\mu(t)\right)\right)$, for $x\in L^2(I)$, $t\in I$. It can be shown that $C$ is a compact, symmetric, positive definite, integral operator with kernel $v$, and thus allows the representation 
\begin{align*}
    Cx(t) = \int_I v(s,t) x(s) \mathrm{d}s,\quad t\in I;
\end{align*}  
see e.g.~\cite{ramsay2005} for more insight. 
Let us for now assume that $\mu=0$. Mercer's theorem~\citep{Mercer1909} then guarantees the existence of an orthonormal basis $\psi_i\in L^2(I)$, for $i\in\mathbb{N}$, and a non-negative sequence $(\lambda_i)_{i\in\mathbb{N}}$, such that the kernel $v$ admits a representation $v(s,t)=\sum_{i=1}^\infty\lambda_i \psi_i(s)\psi_i(t).$ In continuation, we will drop the argument $t$, as the summation and multiplication of functions are naturally defined element-wise. 
Furthermore,  since $C$ is a compact, symmetric operator in $L^2(I)$, it can be represented by a spectral decomposition 
$$
Cx=\sum_{i=1}^\infty\lambda_i\langle\psi_i,x\rangle\psi_i \quad \forall x\in L^2(I),
$$
where $\lambda_i\to 0$, and the convergence of the partial sums is in the operator norm. 
$\psi_i$ is the $i$-the eigenfunction of $C$ corresponding to the eigenvalue $\lambda_i$, i.e., $C\psi_i=\lambda_i\psi_i$ for $i=1,2,\dots$. This further implies that $\lambda_i=\langle C\psi_i,\psi_i\rangle=\langle\mathbb{E}(\langle X,\psi_i\rangle X),\psi_i\rangle=\mathbb{E}(\langle X,\psi_i\rangle ^2)$. 
As $C$ is an integral operator in $L^2(I)$, it is also a trace-class operator, further implying that $\sum_i\lambda_i<\infty$~\citep{ramsay2005}. \\

Let further $X_1,\dots,X_n$ be a functional random sample from $L^2(I)$. Then, the sample mean $\bar{X}$ and the covariance operator $\hat{C}$ are defined as
$$
\bar{X}=\frac{1}{n}\sum_{i=1}^nX_i,\quad \hat{C}(t)=\frac{1}{n}\sum_{i=1}^n(X_i-\bar X)\otimes(X_i-\bar{X}),
$$
where for $f,g\in L^2(I)$ the outer product $f\otimes g:L_2(I)\to L^2(I)$ is the linear operator satisfying  $(f\otimes g)(x)=\langle g,x\rangle f$, for $x\in L^2(I)$.  

For the subset $H\subset\{1,2,\dots,n\}$, $|H|=h$, $n/2\leq h\leq n$ we define 
$$
\bar X_H=\frac{1}{h}\sum_{i\in H}X_i,\quad \hat{C}_H(X)=\frac{1}{h}\sum_{i\in H}(X_i-\bar X)\otimes(X_i-\bar{X}_H),\quad\text{and}\quad \hat{C}_H(X)^\mathrm{NC}=\frac{1}{h}\sum_{i\in H}X_i\otimes X_i,
$$
to be the trimmed sample mean, covariance, and non-centered covariance operators, respectively, calculated at the subset $H$. 
We denote further $\hat{\psi}_{i,H}$ to be the $i$-th eigenfunction of $\hat{C}_H$, with the corresponding 
eigenvalue $\hat{\lambda}_{i,H}$, i.e., $\hat{C}_H\hat{\psi}_{i,H}=\hat{\lambda}_{i,H}\hat{\psi}_{i,H}$, $i=1,2,\dots$. Then, $\hat{C}_H$  has an eigendecomposition $\hat{C}_Hx=\sum_{i=1}^\infty\hat{\lambda}_{i,H}\langle\hat{\psi}_{i,H},x\rangle \hat{\psi}_{i,H}$, where $\sum_{i=1}^\infty\hat{\lambda}_{i,H}<\infty$ for the reasons stated above.

\section{Functional Regularized Minimum Covariance Trace Estimator}\label{sec:MRCT}

\subsection{Tikhonov regularization}
As indicated in Section~\ref{sec:motivation}, the extension of the optimization problem~\eqref{eq:MVT_theorethical} involves defining a standardization of a random function $X\in L^2(I)$. Thus, we proceed by finding the standardized $Y\in L^2(I)$, 
such that it \textit{approximately} solves $C^{1/2}Y=X$, where $C^{1/2}$ is a symmetric squared root of the symmetric operator $C$; $C^{1/2}C^{1/2}X=CX$, for all $X\in L^2(I)$. In general, the problem has no solution in $L_2$ as $C$ is not invertible. We thus employ regularization to the corresponding ill-posed linear problem, particularly via classical Tikhonov regularization with a squared $L_2$-norm penalty. In more formal terms, given the regularization parameter $\alpha>0$, the $\alpha$-standardization of the zero-mean random function $X\in L^2(I)$, with covariance operator $C$, is defined as a solution to the optimization problem
\begin{equation}\label{eq:eq_Tikhonov}
   X_\mathrm{st}^\alpha:=\argmin_{Y\in L^2(I)}\left\{\|C^{1/2}Y-X\|^2+\alpha\|Y\|^2 \right\}. 
\end{equation}
The Tikhonov regularization problem \eqref{eq:eq_Tikhonov} admits a closed-form solution as
$$
X_\mathrm{st}^\alpha=C^{1/2}(C+\alpha I)^{-1}X;
$$
see \cite{cucker_zhou_2007} for more details. By the spectral theorem for compact and self-adjoint operators, we have 
$$
C^{1/2} = \sum_{i=1}^\infty \lambda_i^{1/2} \langle \psi_i , X \rangle \psi_i ~~~ \text{and} ~~~ \left(C + \alpha I \right)^{-1} = \sum_{j=1}^\infty \frac{1}{\lambda_j + \alpha} \langle \psi_j , X \rangle \psi_j,
$$
further giving
\begin{align*}
    X_{\mathrm{st}}^\alpha =&~ ({C}+\alpha I)^{-1}C^{1/2} X =~ \sum_{j=1}^\infty \frac{1}{\lambda_j + \alpha} \langle \psi_j , \sum_{i=1}^\infty \lambda_i^{1/2} \langle \psi_i , X \rangle \psi_i \rangle \psi_j \\
    =&~ \sum_{j=1}^\infty\sum_{i=1}^\infty \frac{\lambda_i^{1/2}}{\lambda_j + \alpha} \langle \psi_i , X \rangle \psi_j \delta_{ij}
    = \sum_{i=1}^\infty \frac{\lambda_i^{1/2}}{\lambda_i + \alpha} \langle \psi_i , X \rangle \psi_i,
\end{align*}
where, $\lambda_j$ and $\psi_j$, $j=1,2,\dots$ are the eigenvalues and the eigenfunctions of $C$, respectively. The covariance operator $C(X_\mathrm{st}^\alpha)$ of $X_\mathrm{st}^{\alpha}$ is then $C(X_\mathrm{st}^\alpha)=C^2(C+\alpha I)^{-2}$, and due to the considerations above it allows the representation $\displaystyle C(X_\mathrm{st}^\alpha)=\sum_{i=1}^\infty \lambda_{i,\mathrm{st}}^\alpha\psi_i\otimes\psi_i$, where $\lambda_{i,\mathrm{st}}^\alpha=\lambda_i^2(\lambda_i+\alpha)^{-2}$, $i\geq 1$; for illustration of the relation between $\alpha$ and $\lambda_{i,\mathrm{st}}^\alpha$ see Figure~\ref{fig:fig1}.  Furthermore, Parseval's identity implies $||X_{\mathrm{st}}^\alpha||^2 = \sum_{i=1}^\infty \frac{\lambda_i}{\left(\lambda_i + \alpha\right)^2} \langle \psi_i, X \rangle^2.$ 

\subsection{Connection to the Reproducing Kernel Hilbert Space}
In an attempt to generalize the concept of the Mahalanobis distance for the functional data in $L_2(I)$, \cite{berrendero2020} proposed to first approximate the random function $X\in L^2(I)$ with the one in the RKHS $\mathcal{H}(C)$, where the kernel in question is, in fact, a covariance-kernel, such that the approximation is "smooth enough". More precisely, for $X\in L^2(I)$ and a regularization parameter $\alpha>0$, define 
 $X_\alpha=\argmin_{f\in\mathcal{H}(K)}\|X-f\|^2+\alpha\|f\|^2_C,$
where 
$$\langle f,g\rangle_C=\sum_{i=1}^\infty\frac{\langle f,\psi_j\rangle\langle g,\psi_j\rangle}{\lambda_j},\quad \|f\|_C^2=\langle f,f\rangle_C.
$$
Proposition 2 in \cite{berrendero2020} shows that $X_\alpha\in \mathcal{H}(C)$ admits a closed form as $X_\alpha=C^{1/2}X_\mathrm{st}^\alpha$. \cite{berrendero2020} then define the $\alpha$-Mahalanobis distance between $X,\,Y\in L^2(I)$ as in Definition \ref{def:alpha_mah}.

\begin{definition}\label{def:alpha_mah}
Given a constant $\alpha>0$, the $\alpha$-Mahalanobis distance (w.r.t. covariance $C$) between $X$ and $Y$ in $L^2(I)$ is defined by
$$
M_{\alpha,C}(X,Y)=\|X_\alpha-Y_\alpha\|_C,
$$
where $X_\alpha$ and $\|\cdot\|_C$ are as defined above. 
\end{definition} 

The $C$-norm of a regularized RKHS approximation $X_\alpha$ of the process $X$ has the form $||X_\alpha||_C^2 = \sum_{i=1}^\infty \frac{\lambda_i}{\left(\lambda_i + \alpha\right)^2} \langle \psi_i, X \rangle^2$. Combining this with the fact that $||X_\alpha||_C^2 = M_{\alpha,C}^2 (X,0)$ yields
\begin{equation} \label{sdnorm}
||X_{\mathrm{st}}^\alpha||^2 = ||X_\alpha||_C^2 = M_{\alpha,C}^2 (X,0) ,
\end{equation}
where, without loss of generality, we assume $\E[X] = 0$. 

Adopting the notion of $\alpha$-Mahalanobis distance as a meaningful dissimilarity measure in this concept, the $\alpha$-standardization process produces a function in $L^2(I)$ whose $L^2$-norm is indeed equal to such dissimilarity between $X$ and its mean. Thus, for solution $X_{\mathrm{st}}^\alpha$ of \eqref{eq:eq_Tikhonov}, the generalization of the optimization problem \eqref{eq:MVT_theorethical} is given by
\begin{align}\label{eq:functional CMT true}
H_0&=\argmin_{\{H\subset\{1,\dots,n\};|H|=h\}}\mathrm{tr}(\hat{C}_{H}^\mathrm{NC}(X_{\mathrm{st}}^\alpha))=\argmin_{\{H\subset\{1,\dots,n\};|H|=h\}}\mathrm{tr}((C+\alpha I)^{-2}C\hat{C}_{H}^\mathrm{NC}(X))\nonumber\\
&=\frac{1}{h}\sum_{i\in H}M_{\alpha,C}^2(X_i,0),
\end{align}
where $\hat{C}_{H}^\mathrm{NC}(X_{\mathrm{st}}^\alpha)$ 
is the non-centered, trimmed sample covariance of the $\alpha$-standardized observations $X_{i,\mathrm{st}}^{\alpha}={C}^{1/2}({C}+\alpha I)^{-1}X_i$, $i\in H$. 
The corresponding estimators $\bar{X}_{H_0}=\frac{1}{h}\sum_{i\in H_0}X_i$ and $k_{H_0}\hat{C}_{H_0}(X)=k_{H_0}\frac{1}{h}\sum_{i\in H_0}(X_i-\Bar{X}_{H_0})\otimes(X_i-\Bar{X}_{H_0})$ are referred to as Minimum Regularized Covariance Trace (MRCT) estimators of the mean and the covariance, respectively. $k_{H_0}$ is the corresponding consistency factor calculated under the assumption of the Gaussianity of the underlying process. In continuation, for simplicity of the notation, we will drop the argument $X$ if the (non-centered) covariance is calculated at the original sample.  

\section{Algorithm}\label{sec:Algorithm}
In practice, in the $\alpha$-standardization process,  the true mean and the covariance operator are unknown. Therefore, we tackle the problem by iteratively replacing them with their current robust estimates, thus yielding an implicit equation for obtaining the optimal subset
\begin{align}\label{eq:fixed_point_alpha_2}
H_0&=\argmin_{\{H\subset\{1,\dots,n\};|H|=h\}}\mathrm{tr}(\hat{C}_{H}^\mathrm{NC}(X_{\mathrm{st},H_0}^\alpha))=\argmin_{\{H\subset\{1,\dots,n\};|H|=h\}}\mathrm{tr}((k_{H_0}\hat{C}_{H_0}+\alpha I)^{-2}k_{H_0}\hat{C}_{H_0}\hat{C}_{H}^\mathrm{NC})\nonumber\\
&=\frac{1}{h}\sum_{i\in H}M_{\alpha,\hat{C}_{H_0}}^2(X_i,\bar{X}_{H_0}),
\end{align}
where 
$\displaystyle
\quad\hat{C}_{H}^\mathrm{NC}(X_{\mathrm{st},H_0}^\alpha)=\frac{1}{h}\sum_{i\in H}X_{i,\mathrm{st},H_0}^\alpha\otimes X_{i,\mathrm{st},H_0}^\alpha,$ 
for $\displaystyle X_{i,\mathrm{st},H_0}^{\alpha}=(k_{H_0}\hat{C}_{H_0})^{1/2}(k_{H_0}\hat{C}_{H_0}+\alpha I)^{-1}(X_i-\bar{X}_{H_0})$, $i\in H$.  $k_{H_0}$ is the consistency factor of $\hat{C}_{H_0}$ determined under the assumption of Gaussianity. In continuation, in order to emphasize on the dependency of $X_{i,\mathrm{st},H}^{\alpha}$ on $k$, we write
$$
X_{i,\mathrm{st},H_0}^{\alpha,k}:=\sqrt{k}\hat{C}_{H_0}^{1/2}(k\hat{C}_{H_0}+\alpha I)^{-1}(X_i-\bar{X}_{H_0}).
$$
Equation~\eqref{eq:fixed_point_alpha_2} thus implies that we search for the $h$-subset $H_0$ used in the estimation of the covariance operator in the set of fixed points of a function $$
f(H_0)=\argmin_{\{H\subset\{1,\dots,n\};|H|=h\}}\mathrm{tr}(\hat{C}_{H}^\mathrm{NC}(X_{\mathrm{st},H_0}^{\alpha,k_{H_0}})).
$$
Lemma~\ref{lem:fixed_point} states that $f$ indeed has at least one fixed point.

\begin{lemma}\label{lem:fixed_point}
Let $\mathcal{H}\subseteq 2^{\{1,\dots,n\}}$ be the set containing all $h$-subsets of $\{1,\dots,n\}$ and define $f:\mathcal{H}\to\mathcal{H}$ with $\displaystyle f(H_0)=\argmin_{\{H\subset\{1,\dots,n\};|H|=h\}}\mathrm{tr}(\hat{C}_{H}^\mathrm{NC}(X_{\mathrm{st},H_0}^\alpha))$. Then $f$ has at least one fixed point. 
\end{lemma}

To estimate the scaling factor $k_{H_0}$ of $\hat{C}_{H_0}$, we employ the strategy used in~\cite{rousseeuw1984}, thus matching the median of the obtained squared $\alpha$-Mahalanobis distances with the median of the corresponding limiting distribution, under the assumption of Gaussianity. For that purpose, we use the following two results. 

\begin{proposition}\label{prop:prop_1}
Let $X\in L^2(I)$, and let $k_{H_0}\hat{C}_{H_0}$ and $\bar{X}_{H_0}$,  $H_0=H_0(n)\to \infty$, as $n\to\infty$ be strongly consistent estimators of the covariance operator $C$ and the mean $\mu=0$, respectively. Then, for $\alpha>0$,  $\|X_{\mathrm{st},H_0}^{\alpha,k_{H_0}}\|\to_{a.s.} \|X_{\mathrm{st}}^\alpha\|$.
\end{proposition}

\begin{cor}\label{cor:cor_2}
Let $X\in L^2(I)$ be a Gaussian random process and let $k_{H_0}\hat{C}_{H_0}$ and $\bar{X}_{H_0}$,  $H_0=H_0(n)\to \infty$, as $n\to\infty$ be strongly consistent estimators of the covariance operator $C$ and the mean $\mu=0$, respectively. Then, for $\alpha>0$,  $\|X_{\mathrm{st},H_0}^{\alpha,k_{H_0}}\|$ converges in distribution to $ \sum_{i=1}^\infty\frac{\lambda_i^2}{(\lambda_i+\alpha)^2}Y_i$, where $Y_i$, $i=1,2\dots$ are independent $\chi^2(1)$ random variables and $\lambda_i$, $i=1,2\dots$ are eigenvalues of $C$.
\end{cor}

Corollary~\ref{cor:cor_2} then implies that $k_{H_0}$ could be estimated by matching the sample median of the squared $\alpha$-Mahalanobis distances (i.e.~$\|X_{i,\mathrm{st},H_0}^{k_{H_0},\alpha}\|^2$, $i=1,\dots,n$) with the estimate of the theoretical median of the weighted sum of independent $\chi^2(1)$ random variables, as described in the result. However, as eigenvalues ${\lambda}_{i}$, $i\geq 1$ of the true covariance are unknown, we estimate them with the eigenvalues $k_{H_0}\hat{\lambda}_{i,H_0}$, $i\geq 1$ of the current robust estimate $k_{H_0}\hat{C}_{H_0}$ of the covariance operator and
 choose the scaling parameter $k$ such that 
\begin{align}\label{eq:k}
\mathrm{med}\left\{\|X_{i,\mathrm{st},H_0}^{\alpha,k}\|^2:i=1,\dots,n\right\}&=\mathrm{med}\left(\sum_{i=1}^\infty\frac{k^2\hat{\lambda}_{i,H_0}^2}{(k\hat{\lambda}_{i,H_0}+\alpha)^2}\chi^2(1)\right)\nonumber\\ 
\iff k^{-1}\mathrm{med}\left\{\|X_{i,\mathrm{st},H_0}^{\alpha/k,1}\|^2:i=1,\dots,n\right\}&=\mathrm{med}\left(\sum_{i=1}^\infty\frac{\hat{\lambda}_{i,H_0}^2}{(\hat{\lambda}_{i,H_0}+\alpha/k)^2}\chi^2(1)\right)\nonumber\\
\iff k&=\frac{\mathrm{med}\left\{\|X_{i,\mathrm{st},H_0}^{\alpha/k,1}\|^2:i=1,\dots,n\right\}}{\mathrm{med}\left(\sum_{i=1}^\infty\frac{\hat{\lambda}_{i,H_0}^2}{(\hat{\lambda}_{i,H_0}+\alpha/k)^2}\chi^2(1)\right)}.
\end{align}
Thus, for $H_0\subset \{1,\dots,n\}$ we determine $k_{H_0}$ by solving the implicit Equation~\eqref{eq:k}.

The strategy described above for, given $\alpha>0$, finding the solution to~\eqref{eq:fixed_point_alpha_2} is presented in Algorithm~\ref{alg:Functional}, to which we refer to as Minimum Regularized Covariance Trace (MRCT) algorithm.

\begin{algorithm}[H]
\caption{MRCT algorithm}\label{alg:Functional}
\SetKwInOut{Input}{Input}
        \Input{Sample $\{{X}_1,\dots, {X}_n\}$, an initial subset $H_1\subset\{1,\dots,n\}$ with $|H_1|=h$, regularization parameter $\alpha>0$, and the tolerance level $\varepsilon_k>0$;}
        \BlankLine
	\SetKwRepeat{Do}{do}{while}
	\Do{$H_0\neq H_1$}{
	$H_0\leftarrow H_1$;\\
	$\hat{{C}}_{H_0}=\frac{1}{h}\sum_{i\in H_0}({X}_i-\bar{{X}}_{H_0})\otimes({X}_i-\bar{{X}}_{H_0})=\sum_{i=1}^{\infty}{\hat\lambda}_{i,H_0}\hat{\psi}_{i,H_0}\otimes\hat{\psi}_{i,H_0}$;\\
	$k_1\leftarrow 1$;\\
	\Do{$(k_1-k_0)^2\geq\varepsilon_k$}{
	    $k_0\leftarrow k_1$;\\
	    Calculate $H_0$-$\alpha/k_0$-standardized observations $X_{i,\mathrm{st},H_0}^{\alpha/k_0,1}=\hat{C}_{H_0}^{-1/2}(\hat{C}_{H_0}+\alpha/k_0 I)^{-1}(X_i-\bar{X}_{H_0})$;\\ 
	    Calculate $d_{i,H_0,k_0}^2=\|X_{i,\mathrm{st},H_0}^{\alpha/k_0,1}\|^2$;\\
	   $k_1\leftarrow{\mathrm{med}\left\{d_{i,H_0,k_0}^2:i=1,\dots,n\right\}}/{\mathrm{med}\left(\sum_{i=1}^\infty\frac{\hat{\lambda}_{i,H_0}^2}{(\hat{\lambda}_{i,H_0}+\alpha/k_0)^2}\chi^2(1)\right)}$;\\
	}
    Order $d_{(i_1,H_0,k_1)}\leq\dots\leq d_{(i_n,H_0,k_1)}$;\\
    Set $H_1\leftarrow\{i_1,\dots,i_h\}$;\\
    }
    \SetKwInOut{Output}{Output}
	\Output{Subset $H_1$, robust squared $\alpha$-Mahalanobis distances $(k_1^{-1}d_{(1,H_0,k_1)}^2,\dots k_1^{-1}d_{(n,H_0,k_1)}^2)$}
\end{algorithm}

The output of the MRCT algorithm is a (possibly not unique) solution to the implicit Equation~\eqref{eq:fixed_point_alpha_2}. 
Therefore, the MRCT algorithm potentially yields multiple  outcomes based on different initial subsets. These then  provide a range of options for selecting the optimal robust covariance and $\alpha$-Mahalanobis distance estimators. In the following, we explore methods for effectively choosing among the candidates generated by the algorithm.
\begin{itemize}
    \item[i)] Given the initial subset $H_0$, Algorithm~\ref{alg:Functional} returns subset $H_1$ for which 
    $$
    \min_{\{H\subset\{1,\dots,n\}:|H|=h\}}\mathrm{tr}(\hat{C}_{H}^\mathrm{NC}(X_{\mathrm{st},H_1}^{\alpha,k_1}))=\mathrm{tr}(\hat{C}_{H_1}^\mathrm{NC}(X_{\mathrm{st},H_1}^{\alpha,k_1})).
    $$
    We then search for a subset $H_1$ satisfying Equation~\eqref{eq:fixed_point_alpha_2} that also minimizes $\mathrm{tr}(\hat{C}_{H_1}^\mathrm{NC}(X_{\mathrm{st},H_1}^{\alpha,k_1}))$, i.e. denoting $\mathcal{X}=\{H_1:|H_1|=h: \min_{\{H\subset\{1,\dots,n\}:|H|=h\}}\mathrm{tr}(\hat{C}_{H}^\mathrm{NC}(X_{\mathrm{st},H_1}^{\alpha,k_1}))=\mathrm{tr}(\hat{C}_{H_1}^\mathrm{NC}(X_{\mathrm{st},H_1}^{\alpha,k_1}))\}$, the optimal subset $H_\mathrm{opt}$ is a solution to the optimization problem:
    \begin{equation}\label{eq:optimal_H}
            H_\mathrm{opt}=\argmin_{H\in\mathcal{X}}\mathrm{tr}(\hat{C}^\mathrm{NC}_{H_1}(X_{\mathrm{st},H_1}^{\alpha,k_1})).
    \end{equation}
    As the trace of the covariance of the $\alpha$-standardized functions calculated at the $H$-subset equals the sum of the $\alpha$-Mahalanobis distances of the corresponding original observations, $H_\mathrm{opt}$ in \eqref{eq:optimal_H} is essentially taken to be the subset whose observations have the smallest $\alpha$-Mahalanobis distance, w.r.t. the corresponding covariance. The idea is indeed intuitive since the $\alpha$-Mahalanobis distances should be small for regular observations. The discussed criterion could be modified by taking the sum of the first $q\%$ of $\alpha$-Mahalanobis distances, for some $q$, where the considered percentage must not be too high, in order not to include outlying observations (e.g.~$q = 75\%$). 

\item[ii)] An alternative (data-driven) strategy for choosing the optimal solution to \eqref{eq:fixed_point_alpha_2} would focus on searching for such a subset $H\in\mathcal{X}$ for which the obtained $\alpha$-Mahalanobis distances group into two clusters corresponding to non-outlying and outlying observations. If the outliers would be known, one would prefer choosing such a data subset for which the $\alpha$-Mahalanobis distances $M_{\alpha,k_{H_0}\hat{C}_{H_0}}(X_i,\bar{X}_{H_0})$, $i=1,\dots,n$, when grouped based on outlyingness status, have a maximal ratio of the between- to within-cluster variability.  As the class labels are unknown, an approach is to consider an unsupervised classification instead, where the objective is the performance measure of the chosen clustering algorithm, e.g. the ratio of the between- to within-cluster variability (see Example \ref{sec:Examples:2}), or e.g. the Davis-Bouldin index \citep{davies1979}. For such purpose, one could also e.g. use the squared excess kurtosis, i.e. the squared standardized fourth moment of the $\alpha$-Mahalanobis distances. The kurtosis of the univariate projections is widely used as a projection index in independent component analysis (ICA) and clustering; see e.g.~ \cite{Radojicic2021}, while the use of kurtosis as a projection index in outlier detection is discussed in more detail e.g.~in \cite{pena2001}.   If outlying curves are expected, higher squared excess kurtosis could indicate a better separation between regular and irregular observations. 
\end{itemize}

We emphasize that selecting the appropriate value of $\alpha$ can significantly reduce the sensitivity of the method to the initial approximation. Consequently, it is observed that the choice of $H_{\mathrm{opt}}\in\mathcal{X}$ has minimal influence on the final outcomes, including the covariance estimate and outlier detection. For more details see Appendix \ref{sec:appendix1}.

\subsection{Automated selection of the regularization parameter $\alpha$ and the subset size $h$}
\label{sec:regparam}
In the following, we propose an automated approach for selecting the most suitable regularization parameter  $\alpha$. We commence by demonstrating the degree of continuity exhibited by Algorithm \ref{alg:Functional} in relation to $\alpha$.

\citet[Proposition~9]{berrendero2020} show that for a sequence $\alpha_n>0$, with $\alpha_n\to\alpha$ as $n\to \infty$, 
 $\|X_{\alpha_n}^{st}\|\to \|X_{\alpha}^{st}\|$. The subsequent proposition establishes the validity of the statement even when the standardization of $X$ is performed using the sample covariance operator, which is computed based on a specific subset $H_0\subset \{1,\dots,n\}$. 

\begin{proposition}\label{prop:prop_3}
Let $X\in L^2(I)$, and let $k_{H_0}\hat{C}_{H_0}$ and $\bar{X}_{H_0}$, $H_0=H_0(n)\to \infty$, as $n\to\infty$ be strongly consistent estimators of the covariance operator $C$ and mean $\mu=0$, respectively. Then, for a sequence  $\alpha_n>0$,  $\alpha_n\to \alpha>0$,  $\|X_{\mathrm{st},H_0}^{\alpha_n,k_{H_0}}\|\to_{a.s.} \|X_{\mathrm{st}}^\alpha\|$.
\end{proposition}

Proposition~\ref{prop:prop_2} implies that the crucial step in the Algorithm~\ref{alg:Functional} is continuous w.r.t. $\alpha$. More precisely, if $\alpha_1$ and $\alpha_2$ are close enough (so that the ordering of the Mahalanobis distances in \texttt{STEP 8} of Algorithm~\ref{alg:Functional} is preserved; see Proposition~\ref{prop:prop_2}), given the same initial estimate $H_0$, using either $\alpha_1$ or $\alpha_2$ would produce the same updated subset $H_1$. 
\begin{proposition}\label{prop:prop_2}
Let $X\in L^2(I)$, and let $k_{H_0}\hat{C}_{H_0}$ and $\bar{X}_{H_0}$ be the $H_0$-subset sample  estimators of the covariance operator $C$ and mean $\mu=0$, respectively. Then, for every $\varepsilon>0$, there exists $\delta>0$, such that for  $\alpha_1,\,\alpha_2>0$,  $|\alpha_1-\alpha_2|<\delta$,  $|\|X_{\mathrm{st},H_0}^{\alpha_1,k_{H_0}}\|-\|X_{\mathrm{st},H_0}^{\alpha_2,k_{H_0}}\||<\varepsilon$.
\end{proposition}
The proof of Proposition \ref{prop:prop_2} follows the reasoning in the proof of Proposition~\ref{prop:prop_3} and is thus omitted. 

\begin{figure}[htbp]
    \centering
    \includegraphics[width=.8\linewidth]{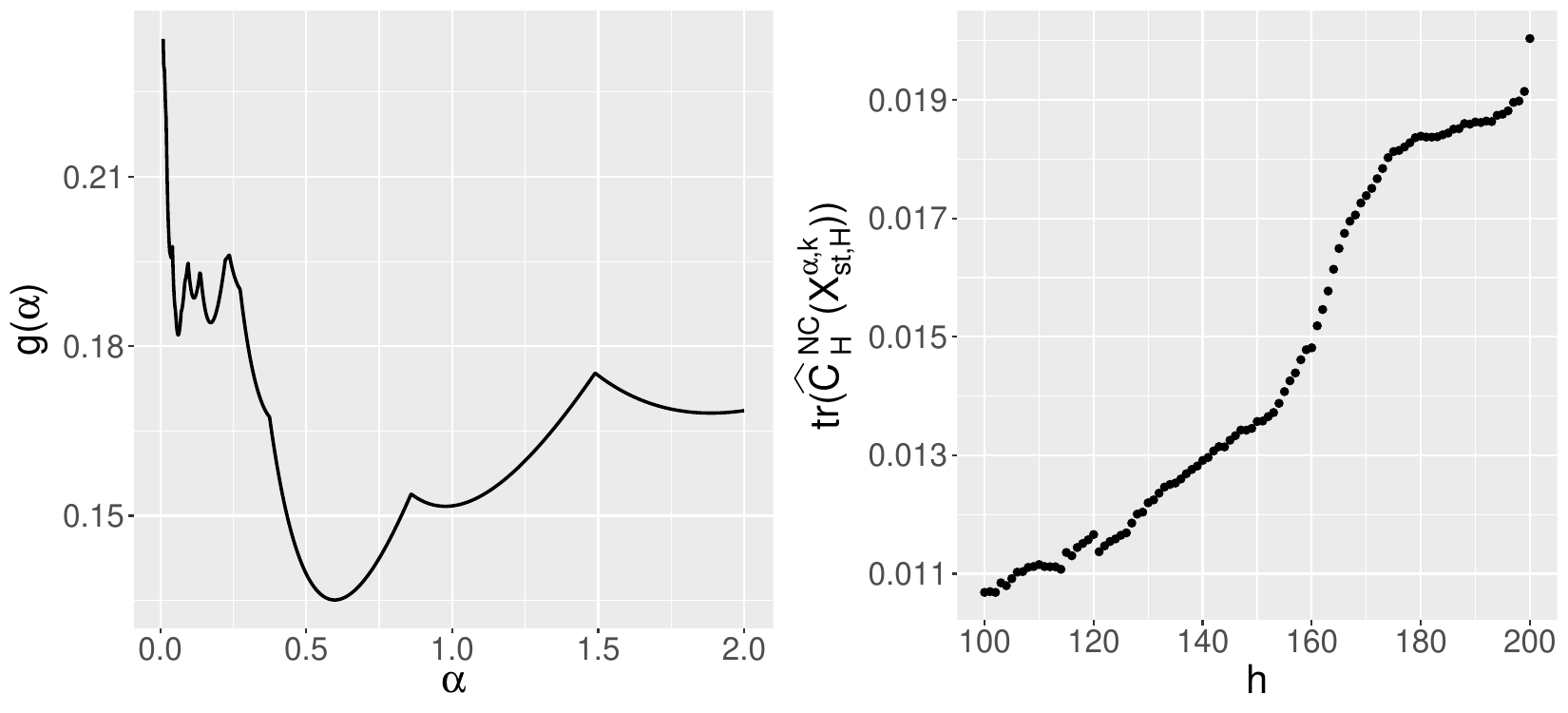}
    \caption{The left plot displays a typical trajectory of the objective for selecting $\alpha$ for Model 1 $(n = 200, p = 100, c = 0.2)$ in the simulation study in Section~\ref{sec:Simulations}. The right plot shows the objective in \eqref{eq:optimal_H} with respect to subset size $n/2 \leq h \leq n$, corresponding to the setting on the left and $\alpha \approx 0.6$. A sudden increase is noticeable around a value of $h=160$, and corresponds to the moment where the first outliers would be included in the subset.
    }
    \label{fig:objective}
\end{figure}

Remark~\ref{rem:rem1} argues that the use of the trace of the covariance of non-standardized observations as an objective results in spherically-biased subsets, where such an obstacle is mitigated by first standardizing the multivariate data so that the obtained covariance is proportional to the identity. As no random function in $L_2(I)$ has an identity covariance, the approach is to choose $\alpha$ such that the eigenvalues of $C(X_{\mathrm{st}}^{\alpha})$ are either as close to $0$ (for the lower part of the spectrum), or as close to some non-zero value, making them also mutually as close together as possible (for the upper part of the spectrum). More precisely, the aim is to select $\alpha$ so that the eigenvalues $\lambda_{i,\mathrm{st}}^{\alpha}$ of $C(X_{\mathrm{st}}^{\alpha})$ cluster into two groups, where the center of one of those groups is $0$, and the joint within-cluster sum of squares is minimal. The approach amounts to minimizing the noise in the data, while at the same time making all the signal components "equally important"; see Figure \ref{fig:fig1} in the Appendix for illustration.  On the sample level, for fixed $\alpha>0$ and the robust estimate of the covariance $k_{H_0}\hat{C}_{H_0}(X_{\mathrm{st},H_0}^{\alpha,k_{H_0}})$, we first calculate $\hat{\lambda}_{i,\mathrm{st},H_0}^\alpha=(k_{H_0}\hat{\lambda}_{i,H_0})^2/(k_{H_0}\hat{\lambda}_{i,H_0}+\alpha)^2$, where $\hat{\lambda}_{i,H_0}$ is the $i$th eigenvalue of $\hat{C}_{H_0}(X_{\mathrm{st},H_0}^{\alpha,k_{H_0}})$, $i=1,\dots,p$. Next, we search for a partition $\{S_1^\alpha,S_2^\alpha\}$, $S_1^\alpha\cup S_2^\alpha=\{1,\dots,p\}$, $S_1^\alpha\cap S_2^\alpha=\emptyset$, such that 
\begin{align*}
\{S_1^\alpha,S_2^\alpha,c_{S_1^\alpha}\}=\argmin_{\{A_1,A_2,c_{A_1}\}: c_{A_1}>0,\,A_1\cup A_2=\{1,\dots,p\},\,A_1\cap A_2=\emptyset }V(A_1;c_{A_1})+V(A_2;0),
\end{align*}
where $V(A_1;c_{A_1})=\sum_{i\in A_1}\left(\hat{\lambda}_{i,\mathrm{st},H_0}^\alpha-c_{A_1}\right)^2$, $V(A_2;0)=\sum_{i\in A_2}(\hat{\lambda}_{i,\mathrm{st},H_0}^\alpha)^2$. Observe first that as $\hat{\lambda}_{1,H_0}\geq \hat{\lambda}_{2,H_0}\geq\cdots $, and due to the monotonicity of $x\mapsto x^2/(x+\alpha)^2$, ordering is preserved for the standardized eigenvalues. $\mathrm{tr}(\hat{C}_{H_0})<\infty$, giving $\sum_{i\geq 1}\hat{\lambda}_{i,H_0}^4\leq \infty$, and further implying $\sum_{i\geq 1}\left(\hat{\lambda}_{i,\mathrm{st},H_0}^\alpha\right)^2<0$ and $\lim_{i\to\infty}\left(\hat{\lambda}_{i,\mathrm{st},H_0}^\alpha\right)^2\to 0$. Thus, for any $c_{A_1}>0$, there exists $i_0\in\mathbb{N}$, such that $(\hat{\lambda}_{i,\mathrm{st},H_0}^\alpha-c_{A_1})^2>(\hat{\lambda}_{i,\mathrm{st},H_0}^\alpha)^2$, implying that the set $S_1$ is finite, and the problem of finding an optimal partition amounts to finding $m_\alpha>0$ such that 
\begin{align*}
m_\alpha=\argmin_{m>0}V(\{1,\dots,m\};\frac{1}{m}\sum_{i=1}^m\hat{\lambda}_{i,\mathrm{st},H_0}^\alpha)+V(\{m+1,m+2,\dots\};0).
\end{align*}
The value of an objective function used for $\alpha$-selection is then 
$$
g(\alpha)=\left (V(\{1,\dots,m_\alpha\};\frac{1}{m}\sum_{i=1}^m\hat{\lambda}_{i,\mathrm{st},H_0}^\alpha)+V(\{m_\alpha+1,m_\alpha+2,\dots\};0)\right)/\left(\frac{1}{m}\sum_{i=1}^m\hat{\lambda}_{i,\mathrm{st},H_0}^\alpha\right)^2,$$ 
and we search for $\alpha\in\argmin g(\alpha)$; see left plot of Figure \ref{fig:objective} for illustration. 
The approach is to optimize $g$ iteratively starting with an initial value of $\alpha_0$.   For an initial alpha, robust estimates of eigenvalues of $C$ are calculated and the value of the objective function $g$ is approximated for each $\alpha$ considered in the optimization. The optimal $\alpha$ found that way is then used for re-estimation of the robust eigenvalues of $C$. The process is repeated until convergence; see  Algorithm \ref{alg:alpha}. In the simulation study (Section~\ref{sec:Simulations}) we use $\alpha_0=0.01$ for $p<n$ (see simulation study in \cite{berrendero2020} for more insight), and $\alpha_0=1$, for $p>n$. However, we noticed that the initial value of $\alpha_0$ has little to no effect on the output of Algorithm~\ref{alg:alpha}. We emphasize that, due to Proposition~\ref{prop:prop_2}, there is no need to run Algorithm~\ref{alg:alpha} on a very dense grid of $\alpha$'s, thus additionally lowering the computational burden. 
\begin{algorithm}[H]
\caption{Algorithm for choosing the regularization parameter $\alpha$}\label{alg:alpha}
\SetKwInOut{Input}{Input}
        \Input{Sample $\{{X}_1,\dots, {X}_n\}$; initial approximation $\alpha_{i_1}$; grid $\{\alpha_1\leq,\alpha_2\leq\cdots\leq\alpha_m\}$ for $\alpha$, over which $g$ is optimized;}
        \SetKwRepeat{Do}{do}{while}
	\Do{$\alpha_0\neq \alpha_{01}$}{
        $\alpha_{0}\leftarrow \alpha_{i_1}$;\\
         For $\alpha_i>0$, run MRCT algorithm \eqref{alg:Functional} and obtain
         robust estimates $\{k_{H_0}\hat{\lambda}_{i,H_0}^{\alpha_{i_1}}:i\geq 1\}$ of eigenvalues of $C$\;
       \For{$i\geq 1$}{
         Calculate $\hat{\lambda}_{i,\mathrm{st},H_0}^{\alpha_{i_1}}=(k_{H_0}\hat{\lambda}_{i,H_0}^{\alpha_{i_1}})^2/(k_{H_0}\hat{\lambda}_{i,H_0}^{\alpha_{i_1}}+\alpha_i)^2$, $i\geq 1$\;
         $m_{\alpha_i}\leftarrow\argmin_{m>0}V(\{1,\dots,m\};\frac{1}{m}\sum_{i=1}^m\hat{\lambda}_{i,\mathrm{st},H_0}^\alpha)+V(\{m+1,m+2\dots\};0)
$\;
     Calculate $g_i=g(\alpha_i;m_{\alpha_i})$\;
     }
     Set $\displaystyle i_1\leftarrow\argmin_{1\leq i\leq m} g_i $
     \;
     }
    \SetKwInOut{Output}{Output}
    \Output{$\alpha_{i_1}$}
\end{algorithm}

Finally, we provide some thoughts about the choice of the subset size $h$. It is clear that the number of contaminated samples should be smaller than $n-h$, and that $h$ should at least be $n/2$ to obtain a fit to the data majority. Therefore, the choice of $h$ is a trade-off between the robustness and efficiency of the estimator; see discussion in \cite{rousseeuw1984} for more details. In practice, the proportion of outliers is, in general, not known in advance, and thus we consider a  data-driven approach for the choice of $h$, as suggested by \cite{rousseeuw2020}: one considers a range of possible values for $h$ to search for significant changes of the computed value of the objective function or the estimate. Big shifts in the norm of the difference between the subsequent covariances estimates, as well as those in the value of the trace of the corresponding covariance based on the consecutive subset sizes, can imply the presence of outliers; see the right plot of Figure \ref{fig:objective} for illustration.

\subsection{Sparsely observed data} \label{sec:sparsedata}
In practice, one does not observe smooth functions, but rather discrete sets of the underlying function values $X_i(t_i):t_i\in\{t_{i,1},\dots,t_{i,p_i}\}$, $i=1, \dots,n$, where $p_{i,j}\in I$, for $i=1, \dots,n$, $j=1,\dots,p_i$. If $p_1=\cdots=p_n=p$ and $t_{i,k}=t_{j,k}$ for all $i,j=1,\dots,n$, and $k=1,\dots,p$, we say that the data is observed on a regular grid. Although there is no precise definition for "dense" functional data, it is conventionally understood that functional data is considered densely sampled when the quantity $p_i$, $i=1,\dots,n$ tends to infinity rapidly enough. If that is the case, then the functional inner products in Algorithm~\ref{alg:Functional} can be approximated by the corresponding integral sums, where the accuracy of the approximation depends on $p$; see e.g.~\cite{ostebee2002} for more details on integral approximations. Given that the data is densely observed, robust estimates of mean and covariance can be approximated empirically at
the observed times $t_i$, $i=1,\dots,p$. Robust estimates of the mean and covariance operator on the entire interval $I$ can then e.g. be obtained by smooth interpolation of the corresponding robust sample estimates \citep{Wang2015ReviewOF}. Smooth interpolation of obtained robust estimates is beyond the scope of this paper; see further e.g.~\cite{Berkey1991,Castro1986} for more insight on FDA for regularly sampled data on a dense grid.

An arguably more common and difficult case is referred to as sparse data. Sparse functional data represents situations where only a small number of irregularly-spaced measurements are available for each observed function. Sparse data may arise due to limitations in data collection or due to the nature of the phenomenon being observed. Analyzing sparse functional data poses unique challenges, as there are fewer data points available to capture the underlying patterns and structures. Therefore, specialized techniques are often employed to effectively handle and extract meaningful information from sparse functional data; see e.g. \cite{ramsay2005,Wahba1990} for a general overview.  
A common approach in FDA when dealing with irregularly observed and sparse data is to assume that the functional data can be represented using a finite set of basis functions. 
We, therefore, express each observed datum $X_i(t),\,i=1,\dots,n$ in a fixed, common basis  $\boldsymbol{\Phi}(t)=(\phi_1(t),\dots,\phi_M(t))'$, $t\in I$ as 
\begin{equation}\label{eq:x_in_bases}
X_i(t)=\sum_{j=1}^M \textbf{C}_{i,j}\Phi_j(t)=\textbf{e}_i'\textbf{C}\boldsymbol{\Phi},\quad i=1,\dots,n,
\end{equation}
where $\textbf{C}$ is a matrix of coefficients in the expansion.  
Then, $\Bar{X}=\frac{1}{n}\sum_{i=1}^n\textbf{e}_i'\textbf{C}\boldsymbol{\Psi}=\frac{1}{n}\textbf{1}'\textbf{C}\boldsymbol{\Psi}$, where $\textbf{1}$ is a vector of ones. 
The number of basis
functions $M$ is chosen individually for a given data set, however, usually in a way that the functional approximations are close to the original observations, with some
smoothing that eliminates most of the noise. The choice of the basis is important, and common choices include polynomial functions, wavelets, Fourier bases, and spline bases, among others; for an overview, see e.g. \cite{ramsay2005}, \cite{kokoszka2017}. Note that $\boldsymbol{\Phi}$ is not necessarily an orthonormal basis, e.g. B-splines \citep[see e.g.][]{Ferraty2007,Eilers1996} is a common choice.
Denote now $\textbf{M}=\langle \boldsymbol{\Phi},\boldsymbol{\Phi}'\rangle$, where 
$\textbf{M}_{i,j}=\langle \phi_i,\phi_j\rangle$, $i,j=1,\dots,M$.  Since the basis functions $\boldsymbol{\phi}$ are linearly independent, $\textbf{M}$ is a regular symmetric matrix. Moreover, for $\textbf{u}\in\mathbb{R}^M$, $\textbf{u}'\textbf{M}\textbf{u}=\|\textbf{u}'\boldsymbol{\Psi}\|^2\geq 0$, where $\|\textbf{u}'\boldsymbol{\Phi}\|^2$ is the squared $L_2$ norm of $\textbf{u}'\boldsymbol{\Phi}\in L_2(I)$.  
Thus, $\textbf{M}^{-1/2}$ is well defined, making $\tilde{\boldsymbol{\Phi}}:=\textbf{M}^{-1/2}\boldsymbol{\Phi}$ orthonormal. Equation \eqref{eq:x_in_bases} can then be rewritten as
\begin{equation}\label{eq:x_in_bases_orth}
X_i(t)=\textbf{e}_i'\textbf{C}\textbf{M}^{1/2}\textbf{M}^{-1/2}\boldsymbol{\Phi}=\textbf{e}_i'\tilde{\textbf{C}}\tilde{\boldsymbol{\Phi}},\quad i=1,\dots,n,
\end{equation}
where $\tilde{\textbf{C}}=\textbf{C}\textbf{M}^{1/2}$ is the matrix of coefficients in the expansion using new, orthonormal bases $\tilde{\boldsymbol{\Phi}}$. This indicates that we can, without loss of generality, assume that the basis functions are orthonormal, i.e. $\textbf{M}=\textbf{I}$, and in the continuation, we proceed by assuming so. Once the data is represented as in~\eqref{eq:x_in_bases_orth}, one might sample the smoothened functions on an arbitrarily dense grid and proceed with an algorithm as in the case where the underlying functions are observed on a regular, dense grid. However, the following proposition indicates that it is not necessary and that we can proceed to work solely with the matrix of coefficients $\textbf{C}$, thus lowering the computational cost and minimizing the approximation error. 

\begin{proposition}\label{prop:Mah_distance_in_a_basis}
 Let $\boldsymbol{\Phi}(t)=(\phi_1(t),\dots,\phi_M(t))'$, $t\in I$  be an orthonormal basis system of functions, and let $X_i$, $i=1,\dots,n$ admit a representation as in~\eqref{eq:x_in_bases}, with $\textbf{C}\in\mathbb{R}^{n\times M}$ being the matrix of coefficients. Then, for any subset $H_0\subseteq\{1,\dots,n\}$, $|H_0|=h,$ the following is true:
 \begin{itemize}
     \item[i)] Let $\hat{\psi}_{i,H_0}$ be the $i$-th eigenfunction of $\hat{C}_{H_0}$, $i=1,\dots,M$. Then, it admits a representation $\hat{\psi}_{i,H_0}(t)=\textbf{u}_i'\boldsymbol{\Phi}(t)$, where $\textbf{u}_i$ is the $i$-th eigenvector of the symmetric matrix $\tilde{\textbf{C}}_{H_0}=\frac{1}{h}\textbf{C}'(\textbf{I}_{n,H_0}-\frac{1}{h}\textbf{J}_{n,H_0})\textbf{C}$; 
     \item[ii)] The $i$-th eigenvalue $\hat{\lambda}_{i,H_0}$, $i=1,\dots,M$, of the sample covariance operator $\hat{C}_{H_0}$ calculated at $H_0$ is also the $i$-th eigenvalue of the symmetric matrix $\tilde{\textbf{C}}_{H_0}$;
     \item[iii)] The squared $\alpha$-Mahalanobis distance of $X_i$ allows for a representation
     $$
      d^2_{i,H_0,k_0} =\|X_{i,\mathrm{st},H_0}^{\alpha/k_0,1}\|^2=(\textbf{e}_i-\frac{1}{h}\mathbf{1}_{H_0})'\textbf{C}\tilde{\textbf{C}}_{H_0}(\tilde{\textbf{C}}_{H_0}+\alpha/k_0\textbf{I}_M)^{-2}\textbf{C}'(\textbf{e}_i-\frac{1}{h}\mathbf{1}_{H_0}), 
     $$
 \end{itemize}
    where $\mathbf{1}_{H_0}\in\mathbb{R}^{n}$ is such that $(\mathbf{1}_{H_0})_i=1$ if $i\in H_0$ and $0$ otherwise, $\textbf{J}_{n,H_0}=\mathbf{1}_{H_0}\mathbf{1}_{H_0}'$, and $\textbf{I}_{n,H_0}=\mathrm{diag}(\mathbf{1}_{n,H_0})\in\mathbb{R}^{n\times n}$ is a diagonal matrix with $\mathbf{1}_{H_0}$ on its diagonal.
\end{proposition}

Using Proposition~\ref{prop:Mah_distance_in_a_basis}, we give the modification of Algorithm~\ref{alg:Functional} in the case where the data are represented in a finite orthonormal basis $\boldsymbol{\Phi}$.

\begin{algorithm}[H] 
\caption{Fixed point algorithm for solving~\eqref{eq:fixed_point_alpha_2} when $\textbf{X}=\textbf{C}\boldsymbol{\Phi}$, for orthonormal basis $\boldsymbol{\Phi}$}\label{alg:Functional_in_basis}
\SetKwInOut{Input}{Input}
        \Input{Orthornormal basis $\boldsymbol{\Phi}=(\phi_1,\dots,\phi_M)$, data matrix $\textbf{X}=({X}_1',\dots, {X}_n')'$, an initial subset $H_1\subset\{1,\dots,n\}$, regularization parameter $\alpha>0$, and tolerance level $\varepsilon_k>0$;}
        \BlankLine
	Express $\textbf{X}$ in a basis $\boldsymbol{\Phi}$ as $\textbf{X}=\textbf{C}\boldsymbol{\Phi}$, where $\textbf{C}\in\mathbb{R}^{n\times M}$ is a matrix of coefficients;\\
	\SetKwRepeat{Do}{do}{while}
	\Do{$H_0\neq H_1$}{
	$H_0\leftarrow H_1$;\\
	$\tilde{\textbf{C}}_{H_0}=\frac{1}{h}\textbf{C}'(\textbf{I}_{n,H_0}-\frac{1}{h}\textbf{J}_{n,H_0})\textbf{C}=\sum_{i=1}^{M}{\hat\lambda}_{i,H_0}\textbf{u}_i\textbf{u}_i'$;\\
	$k_1\leftarrow 1$;\\
	\Do{$(k_1-k_0)^2\geq\varepsilon_k$}{
	    $k_0\leftarrow k_1$;\\
	    Calculate
     $d_{i,H_0,k_0}^2=(\textbf{e}_i-\frac{1}{h}\mathbf{1}_{H_0})'\textbf{C}\tilde{\textbf{C}}_{H_0}(\tilde{\textbf{C}}_{H_0}+\alpha/k_0\textbf{I}_M)^{-2}\textbf{C}'(\textbf{e}_i-\frac{1}{h}\mathbf{1}_{H_0})$;\\
    $ k_1\leftarrow{\mathrm{med}\left\{d_{i,H_0,k_0}^2:i=1,\dots,n\right\}}/{\mathrm{med}\left(\sum_{i=1}^M\frac{\hat{\lambda}_{i,H_0}^2}{(\hat{\lambda}_{i,H_0}+\alpha/k_0)^2}\chi^2(1)\right)}$;\\}
    Order $d_{(i_1,H_0,k_1)}\leq\dots\leq d_{(i_n,H_0,k_1)}$;\\
    Set $H_1\leftarrow\{i_1,\dots,i_h\}$;\\
    }
    \SetKwInOut{Output}{Output}
	\Output{Subset $H_1$, robust squared $\alpha$-Mahalanobis distances $(k_1^{-1}d_{1,H_0,k_1}^2,\dots k_1^{-1}d_{n,H_0,k_1}^2)$}
\end{algorithm}
The adaptation of Algorithm~\ref{alg:alpha} is straightforward and is thus omitted.  
The $\mathrm{rank}(\tilde{\textbf{C}}{H_0})\leq h\leq n$ indicates that the matrix becomes singular when the number of basis functions exceeds the number of observations. However, by adding $\alpha \textbf{I}$ to $\tilde{\textbf{C}}{H_0}$, where $\alpha>0$, the resulting matrix $\tilde{\textbf{C}}_{H_0}+\alpha \textbf{I}$ remains regular. This property allows for the use of an arbitrarily large number of basis functions when expressing the data, thereby reducing the impact of the specific choice of basis. It is important to note that if a small number of basis functions is employed in practice, careful selection of the basis functions becomes crucial to capture most of the data's variability.

\section{Simulation study}\label{sec:Simulations}

This section aims to assess the accuracy of the proposed MRCT algorithm \eqref{alg:Functional} through a simulation study similar to the one conducted in \cite{berrendero2020}. In their work, the authors extended the framework from \cite{arribas-gil2014} to compare various functional outlier detection methods with the newly introduced regularized Mahalanobis distance. 
The selected methods can be classified based on their underlying principles. The first group of outlier detection methods relies on different functional depths, which quantify the outlyingness of curves based on their centrality within a data cloud \citep{fraiman2001}. In this study, we consider the following methods: {Functional boxplots} \citep{sun2011}, which depend on the centered-outward ordering of the Band depth introduced in \cite{pintado2009}; {Depth-based trimming and weighting} \citep{febrero2008}, iterative approaches that iteratively remove outliers from the data until no atypical curves remain; {The adjusted outliergram} \citep{arribas-gil2014}, which combines two depth-based measures into a single index for assessing outlyingness.

The second category of outlier detection methods utilizes functional principal components. In this category, we consider the integrated square error \cite{HYNDMAN2007}, which calculates the integral of the squared difference between a function and its projection onto the first $K$ robust principal components.

The final group of methods is based on the Mahalanobis distance and includes: {The robust Mahalanobis distance}, which treats the curves as multivariate data and calculates this measure based on a robust version of the sample covariance. Additionally, not considered in the previous analysis, we examine the MRCD estimator, which also treats functional data as multivariate; The Mahalanobis distance based on the Reproducing Kernel Hilbert Space (RKHS) \citep{berrendero2020}. However, due to its reliance on the MCD estimator, it can only be applied in the $p<n$ setting. The MRCT algorithm proposed in this paper also falls within this category of methods.

These methods from the different categories collectively form the basis for evaluating the performance of the MRCT algorithm in outlier detection.

In the simulation, we consider three different models that correspond to various outlier settings. These models are designed to capture different scenarios and provide a comprehensive evaluation of the performance of the methods under varying outlier conditions.

\begin{enumerate}
    \item Model 1:
    \begin{itemize}
        \item the main process $X(t) = 30t(1-t)^{1.5} + \varepsilon(t)$,
        \item the contaminated process $X(t) = 30t^{1.5}(1-t) + \varepsilon(t)$,
    \end{itemize}
    where $\varepsilon(t)$ is a Gaussian random process with zero mean and covariance function $\displaystyle\gamma(s,t) = 0.3 \text{exp}\left(\frac{-|s-t|}{0.3}\right)$.
    \item Model 2:
    \begin{itemize}
        \item the main process $X(t) = 4t + \varepsilon(t)$,
        \item the contaminated process $X(t) = 4t + (-1)^u1.8 + (0.02\pi)^{-0.5}\text{exp}\left(\frac{-(t-\mu)^2}{0.02}\right) + \varepsilon(t)$,
    \end{itemize}
    where $\varepsilon(t)$ is a Gaussian random process but this time with covariance function $\gamma(s,t) = \text{exp}(-|s-t|)$. Additionally, $u$ follows a Bernoulli distribution with parameter $0.5$, and $\mu$ is uniformly distributed on the interval $[0.25,0.75]$. 
    \item Model 3:
    \begin{itemize}
        \item the main process $X(t) = 4t + \varepsilon(t)$,
        \item the contaminated process $X(t) = 4t + 2\text{sin}(t(t+\mu)\pi) + \varepsilon(t)$,
    \end{itemize}
    $\varepsilon(t)$ and $u$ are as in Model 2. 
\end{enumerate}

We conduct an analysis for each model by generating a random sample of $n = 200$ observations. These observations are recorded at equidistant time points on the interval $[0,1]$, with two different settings for the number of variables, namely $p = 100$ and $p = 500$. In the context of multivariate statistics, we refer to the $p = 100$ and $p = 500$ settings as low- and high-dimensional, respectively.

\begin{figure}[htp]
\centering
    \subfloat[Model 1]{\includegraphics[width=.33\linewidth]{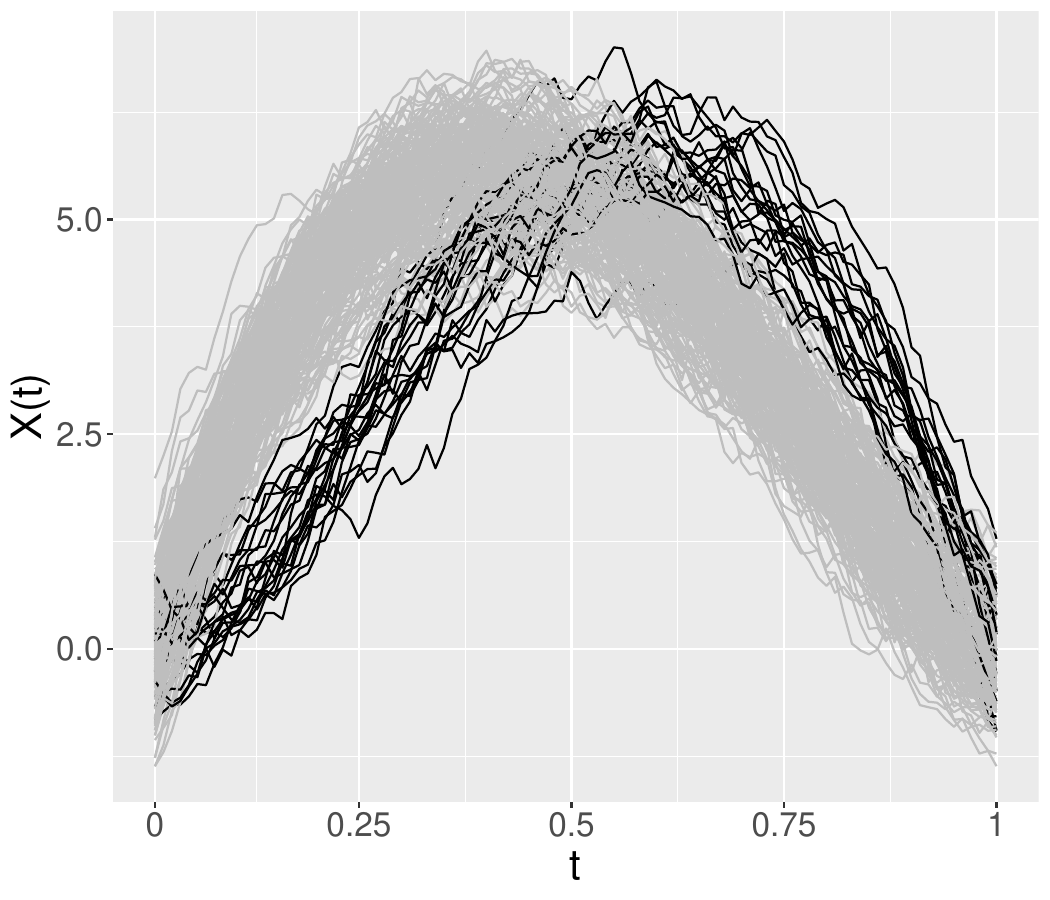}}
    \subfloat[Model 2]{\includegraphics[width=.33\linewidth]{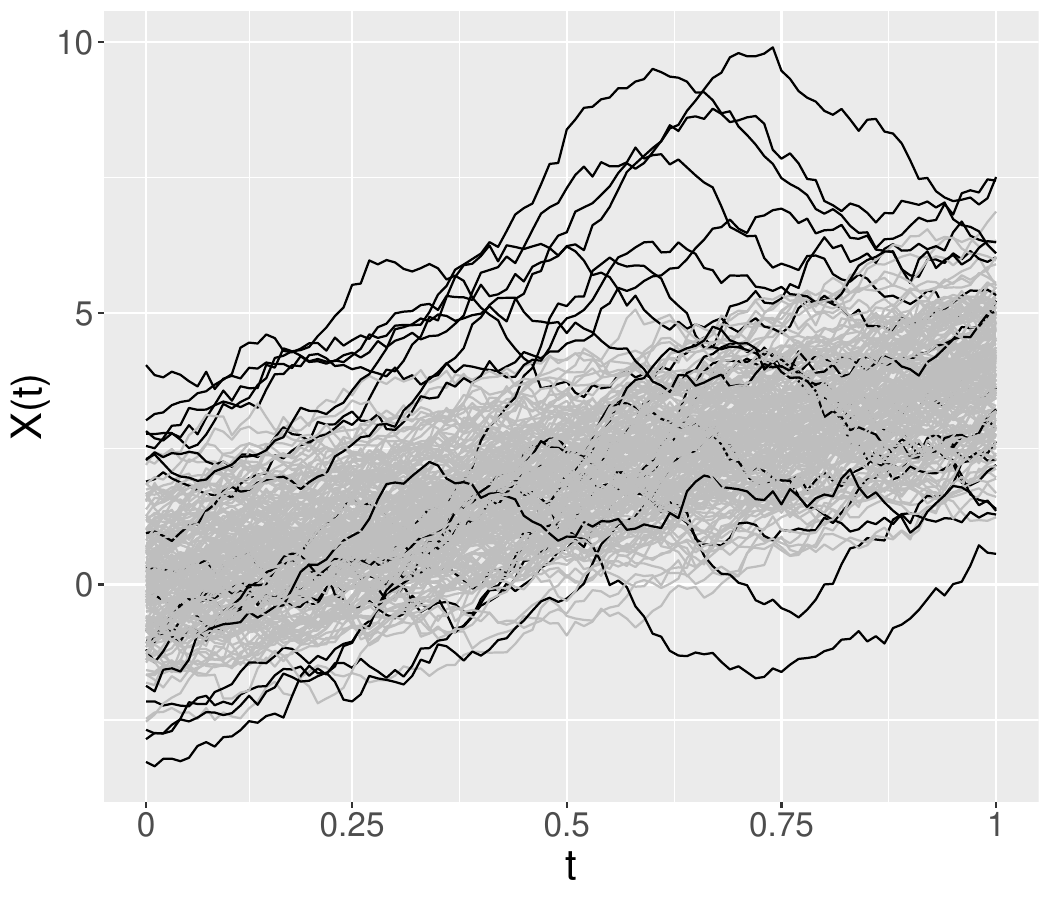}}
    \subfloat[Model 3]{\includegraphics[width=.33\linewidth]{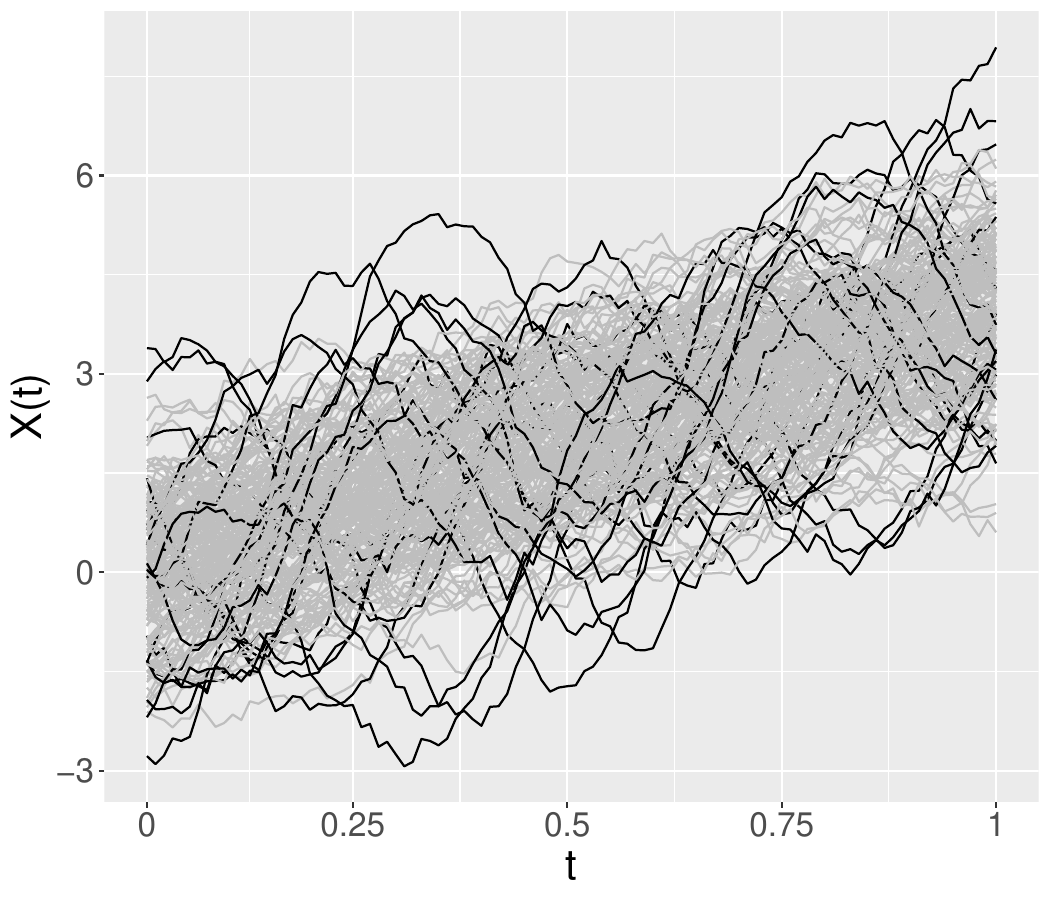}}

    \caption{
    Samples from three models considered in the simulation study. Grey observations represent the main processes, while the black-colored functions indicate the outliers. The contamination rate is $c=0.1$, sample size $n = 200$, and $p = 100$ time points.
    }
    \label{fig:simulationmodels}
\end{figure}

The number of outliers in each setting is determined as $\lfloor n \cdot c \rfloor$, where $c$ represents the contamination rate. In our study, we consider contamination rates of $0$, $0.05$, $0.1$, and $0.2$, allowing us to examine scenarios with no outliers, a small percentage of outliers, and a higher percentage of outliers. Figure \ref{fig:simulationmodels} provides a visual representation of the different main and contaminated processes in the simulation.

To identify the outliers, we compare the obtained robust $\alpha$-Mahalanobis distances from Algorithm \ref{alg:Functional} to the cutoff value corresponding to the estimate of the $99\%$-quantile of the limiting distribution of the robust $\alpha$-Mahalanobis distances, under the assumption of Gaussianity; see Corollary \ref{cor:cor_2}. To estimate the cutoff value, we conduct a Monte Carlo simulation with $2000$ repetitions. 

The performance of all methods is evaluated using three metrics: the true positive rate (TPR), which represents the ratio of correctly classified outliers, the false positive rate (FPR), which represents the ratio of misclassified regular observations, and the F-score calculated as TPR/(TPR+0.5(FPR+FNR)), where FNR corresponds to the ratio of non-identified outliers. Each setting is replicated $50$ times to ensure robustness of the results. For the MRCT method, the subset size $h$ is consistently set to $75\%$ of the data, while the regularization parameter $\alpha$ is determined following the procedure described in Section \ref{sec:regparam}. 

\begin{figure}[htp]
\centering
    \subfloat{\includegraphics[width=1\linewidth]{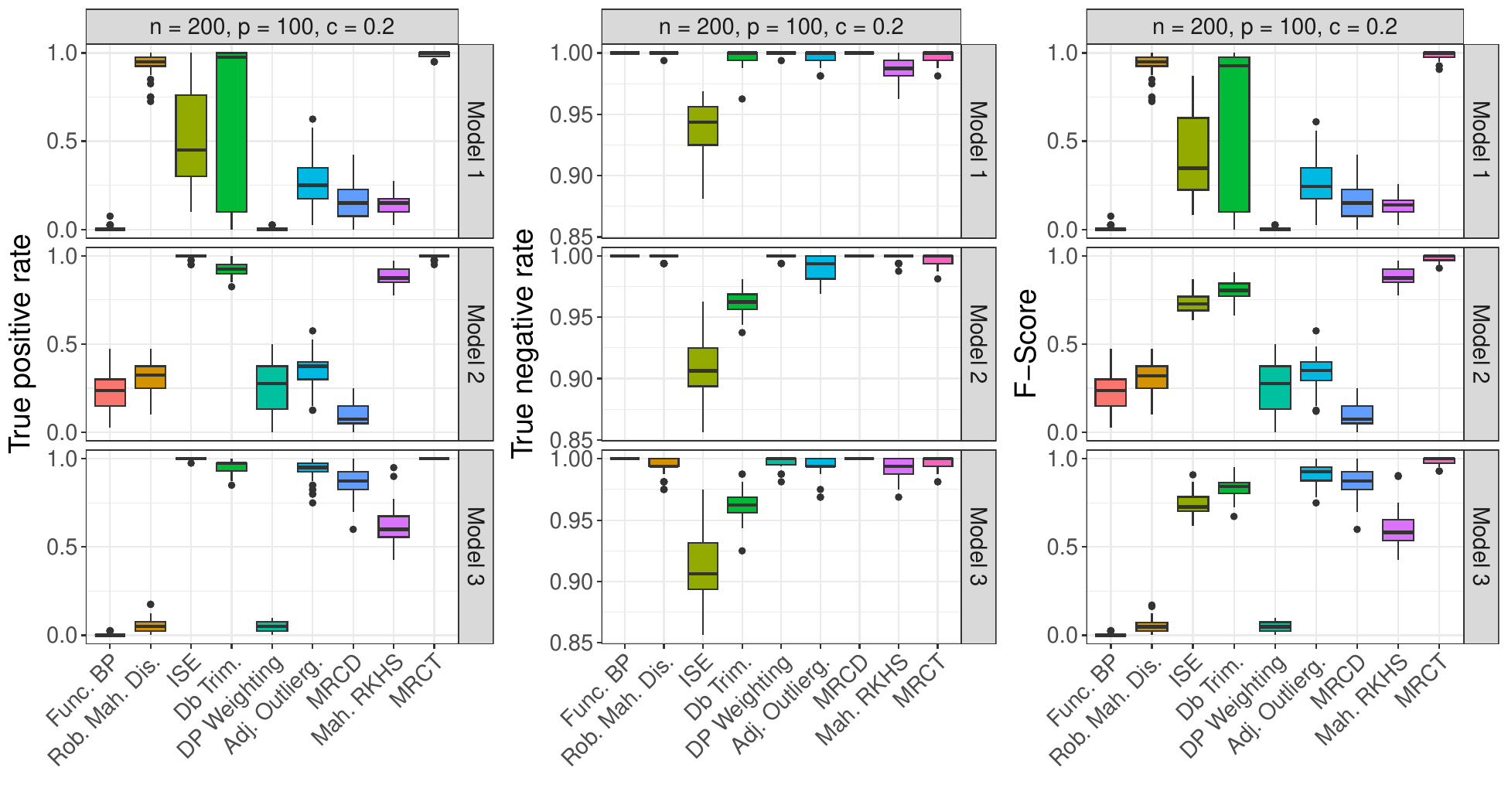}}

    \subfloat{\includegraphics[width=1\linewidth]{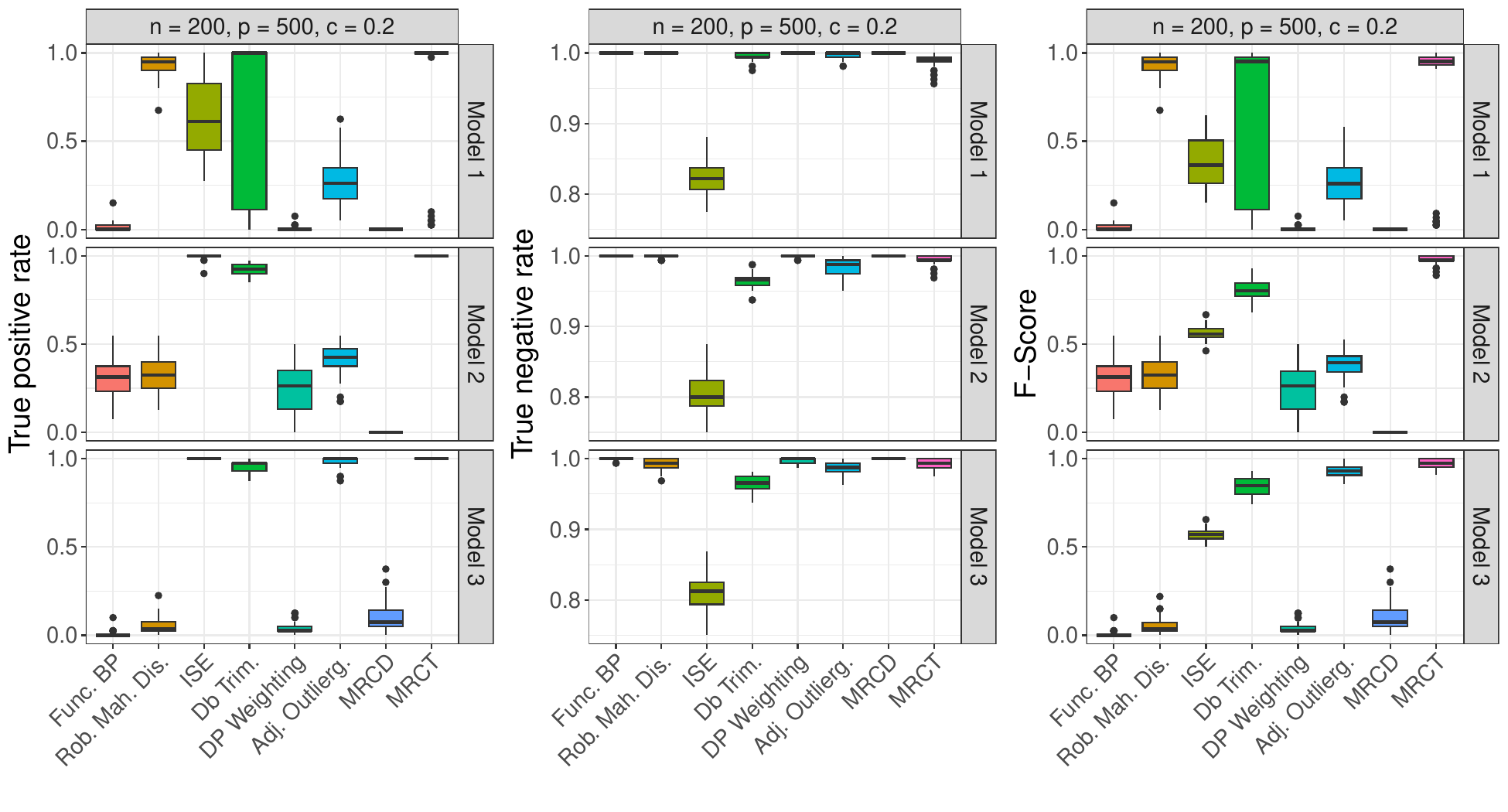}}
    
    \caption{Left to right: boxplots of true positive rate, true negative rate, and F-score for different methods over the three models; $n = 200$, $p = 100, 500$, and $c = 0.2$.}
    \label{fig:simulation_rates}
\end{figure}

Outlier detection rates for all three models are given in Figure \ref{fig:simulation_rates}, where, for the sake of conciseness, we present the results for the contamination rate of $c = 0.2$. The figure includes true positive and true negative rates, as well as the F-score, over 50 replications. Boxplots visualize the rate distributions.

The MRCT algorithm demonstrates good overall performance, often outperforming others in both low- and high-dimensional settings. It achieves high true positive rates, correctly classifying almost all outliers, and maintains desirable true negative rates. Some methods yield higher true negative rates but are rather conservative in the sense that they overlook outliers. The MRCT strikes a good balance, correctly classifying outliers while retaining regular observations.

\begin{figure}[h]
    \centering
    \includegraphics[width=0.8\linewidth]{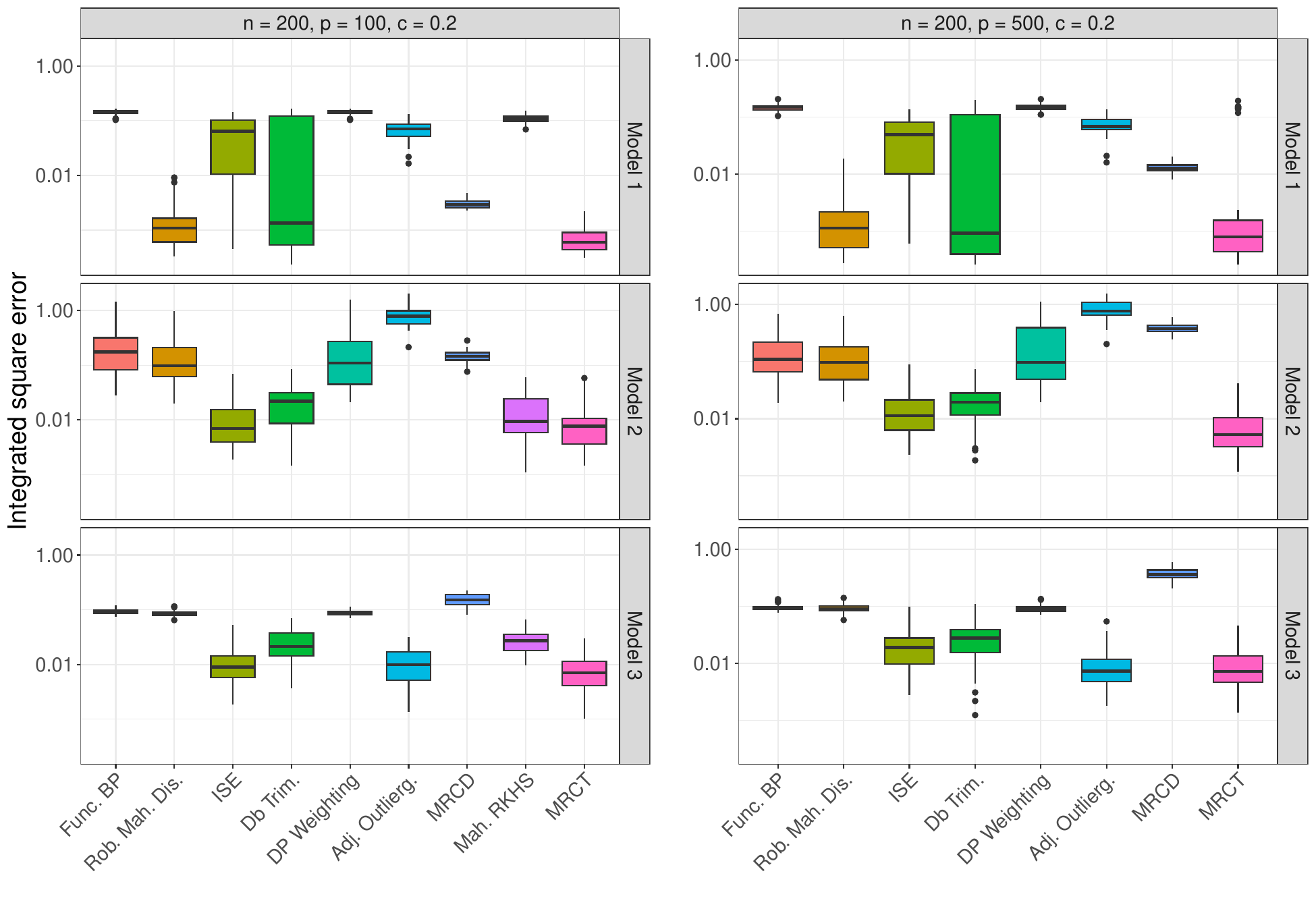}
    \caption{Integrated square error (ISE) of the covariance kernel based on non-outliers and true covariance kernel on a logarithmic scale for $n = 200, p = 100, 500$ and $c = 0.2$.}
    \label{fig:logcoverror}
\end{figure}

In addition to outlier detection, accurate estimation of the data covariance is very important. Therefore, for each replication, we assess the accuracy of the sample estimate $\hat{v}_{reg}$ of the covariance function $\gamma$ of the non-contaminated process using the approximation of an integrated square error (ISE) calculated at the observed time-points; $\text{ISE}=p^{-2}\sum_{i,j=1}^p(\gamma(t_i,t_j)-\hat{v}_{reg}(t_i,t_j))^2$. For each method, $\hat{v}_{reg}$  is a sample covariance estimator calculated at the identified non-outliers.  The results are shown in Figure \ref{fig:logcoverror}.  

\begin{figure}[h]
    \centering
    \includegraphics[width=0.6\linewidth]{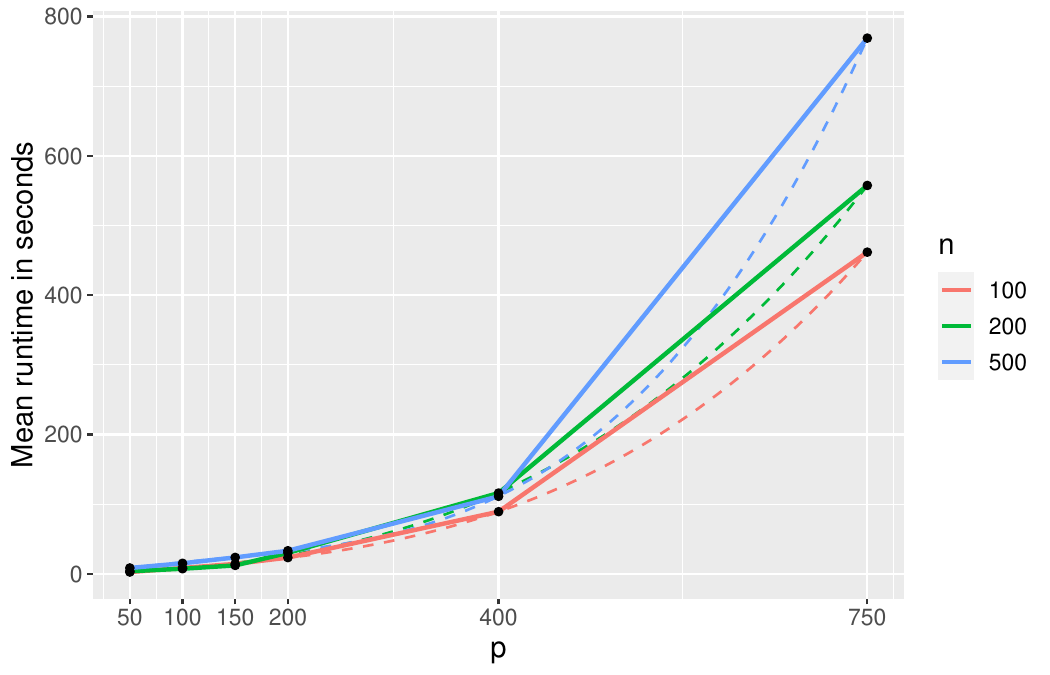}
    \caption{Mean runtime of MRCT algorithm in seconds for Model 2, for different sample sizes $n$, as a function of the number of observations. Dotted lines represent the fitted cubic curve to the mean runtime as a function of $p$.}
    \label{fig:runtime}
\end{figure}

The computational effort is another important aspect of an algorithm. For a fixed iteration of Algorithm \ref{alg:Functional}, in \texttt{STEP 3}, the cost of computing $\hat{C}_{H_0}$  is $O(p^2*h)$, while the cost of computing the corresponding eigendecomposition is $O(p^3)$. Further, for a fixed iteration of the inner loop, in \texttt{STEP 7}, the cost of calculating standardized observations is the cost of matrix-vector multiplication, i.e.~$O(p^2)$, as the eigendecomposition of $\hat{C}_{H_0}$ is precalculated, thus simplifying the calculation of the corresponding matrix square root and inverse.  The simulation study indicated that both the inner and the outer loops converge in just a few iterations, thus suggesting that the cost of Algorithm \ref{alg:Functional}, depending on a number of observed time points $p$ and the subset size $h$ is $O(\max\{h*p^2,p^3\})$. Figure \ref{fig:runtime} depicts the average runtime of Algorithm \ref{alg:Functional} (in seconds), as a function of a number of observed time points $p=50,\,100,\,150,\,200,\,400,\,750$, for the sample size $n=100,\, 200,\,500$. Dotted lines correspond to the cubic function fitted to the average runtime as a function of $p$.

\section{Real data examples}\label{sec:Examples}

\subsection{Sea surface temperatures}

The algorithm is applied to two real-world datasets, starting with an example focusing on the El Niño-Southern Oscillation (ENSO) phenomenon in the equatorial Pacific Ocean. ENSO is a recurring climate pattern characterized by two well-known phases called \textit{El Ni\~no} and \textit{La Ni\~na}. These events involve unusual warming of the ocean surface, which can have global impacts on weather conditions \citep{elnino}. One classification method for identifying these events is the Oceanic \text{El Ni\~no} Index (ONI), which utilizes 3-month running means of sea surface temperature (SST) in specific regions of the Pacific Ocean. An indicator of an ENSO event is when five consecutive means are above or below a threshold. The relevant region for the analysis is \text{Niño} 1+2,  located near the west coast of South America. Weekly temperature data, available at \url{https://www.cpc.ncep.noaa.gov/data/indices/},  from March 1982 to February 2023 is considered, with each yearly episode consisting of the initial observation in March and the following 52 weeks, resulting in a $41\times 53$ data matrix.

\begin{figure}[h]
    \centering
    \includegraphics[width=\linewidth]{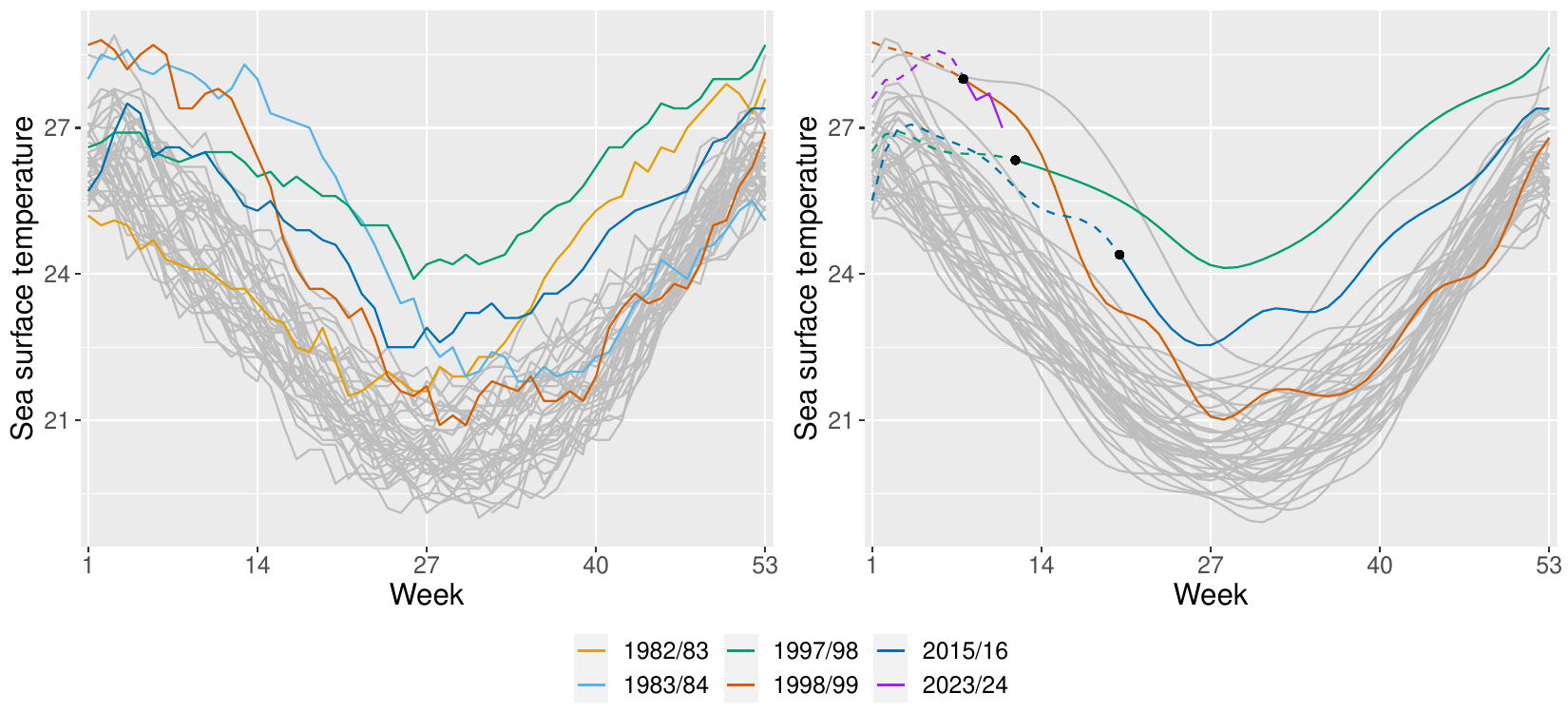}
    \caption{SST of the Ni\~no 1+2 region in the Pacific Ocean. The dataset spans from 1982 to the present. In the left plot, regular observations are depicted in grey, while outliers are represented by colored markers. The right plot showcases the outliers of the smoothed SST data, determined through forecasting based on the available historical information. The black dots indicate the specific points in time when El Ni\~no events were detected solely using past information. 
    }
    \label{fig:ninooutliers}
\end{figure}

The objective of this analysis is to identify atypical behavior in the yearly sea surface temperature, indicating the presence of an El Ni\~no event. The regularization parameter, denoted as $\alpha$, is automatically selected using Algorithm \ref{alg:alpha} and set to $\alpha = 2$. The subset size $h$ is fixed at $\lfloor 0.75 n  \rfloor$. The identified outliers are depicted in Figure \ref{fig:ninooutliers}, and correspond to the seasons of $1982/83$, $1983/84$, $1997/98$, $1998/99$, and $2015/16$. These seasons are associated with some of the strongest El Ni\~no events in the years $1982/83$, $1997/98$, and $2015/16$, and the outliers have also been detected using other methods, as referenced e.g. in  \cite{huang2019} or \cite{suhaila2021}.

The prediction of such events holds significant relevance in the agricultural sector. Therefore, it is of interest to investigate whether these events can be forecasted using information from preceding years, either at the beginning or after a few months into the episodes. In this study, we explore this notion by initially considering only the temperatures of the first 8 weeks, approximately corresponding to March and April, and subsequently extending the temperature values by incorporating data from the following weeks, making it in total $46$ data sets in which outlier detection is performed. As no information was available prior to $1982$, we specifically focus on the more recent outlier episodes of $1997/98$, $1998/99$, and $2015/16$. In cases where the data is sparse due to a limited number of available weeks, we apply data smoothing techniques as described in Section \ref{sec:sparsedata} using Algorithm \ref{alg:Functional_in_basis}. We utilize a B-spline basis, where the number of basis functions depends on the number of weeks but is capped at a maximum of 15.

The findings are presented in Figure \ref{fig:ninooutliers}, showcasing the results obtained from analyzing the smoothed data. For the episode spanning from $1997$ to $1998$, the outlying behavior emerges at the beginning of June, as denoted by the black dot. Subsequently, the observations consistently exhibit outlying characteristics, leading to the different linetype of the corresponding curve from that point onward. By initially considering the data comprising the temperatures of the first 8 weeks, the observations from the following year, $1998/99$, are promptly identified as outliers due to their continuation of the trend of above-average sea surface temperatures. As the year progresses, the El Ni\~no event weakens, evident from the declining SST values. The subsequent El Ni\~no event in $2015/16$ exhibits distinct patterns, suggesting differences in its underlying dynamics. Consequently, this observation is only identified as an outlier starting from mid-July.

Furthermore, an investigation of the current season, $2023/2024$, was also conducted. At the time of analysis, only 11 weeks of information until mid-May were available. By performing a similar forecasting study using the first 8 to 11 weeks of data from all available years, indications suggest the onset of the next El Ni\~no event. This is represented by the purple curve in Figure \ref{fig:ninooutliers}, displaying abnormally high sea surface temperatures around March and April. Similar to the $1998$ El Ni\~no event, it is already detected from the initial stages.

\subsection{EPXMA spectra}\label{sec:Examples:2}

\begin{figure}[h]
    \centering
    \includegraphics[width=\linewidth]{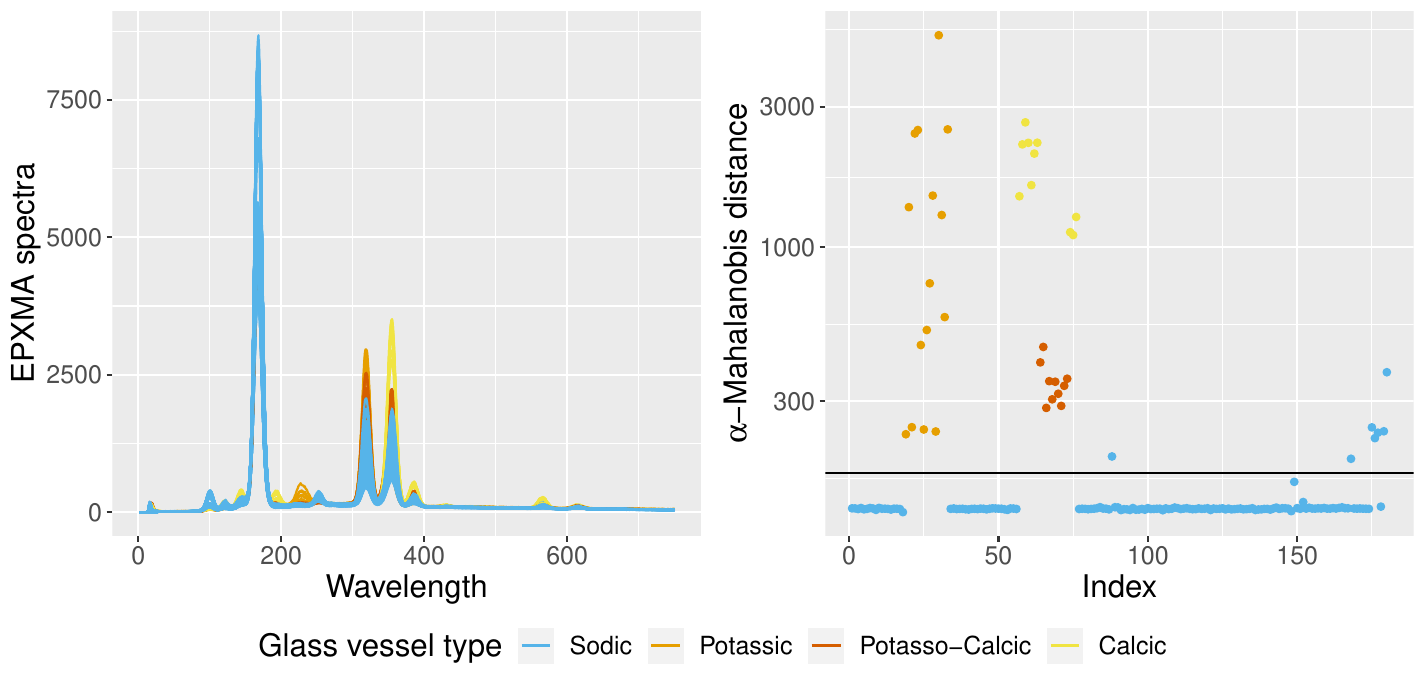}
    \caption{Left: Spectra of different glass vessel types over several wavelengths. Right: $\alpha$-Mahalanobis distance of non-outliers on a logarithmic scale. The horizontal solid line indicates the theoretical cut-off value.}
    \label{fig:epxma}
\end{figure}

The following datasets comprise electron probe X-ray microanalysis (EPXMA) spectra, specifically spectra of various glass vessel types. These data were obtained during an analysis conducted in the late 1990s at the University of Antwerp, with the objective of studying the production and trade relationships among different glass manufacturers \citep{janssens1998}. The glass vessels, originating from the 16th and 17th centuries, were subjected to archaeological analysis. A total of 180 observations were recorded for 1920 wavelengths. Our focus is on the first 750 timepoints, as they encapsulate the primary variability within the data. The data comprises four distinct types of glass vessels. The primary group, referred to as the "sodic" group, encompasses 145 observations. The remaining three groups are denoted as "potassic," "potasso-calcic," and "calcic," consisting of 15, 10, and 10 measurements, respectively. Further details regarding the different glass types can be found in \cite{lemberge2000}. The corresponding curves are depicted in the left plot of Figure \ref{fig:epxma}. Notably, the potassic and calcic observations exhibit deviant shapes from the majority at approximately 200, in addition to having higher amplitudes within the 300 to 350 wavelength range. On the other hand, the potasso-calcic group does not exhibit apparent outlying characteristics.

For this analysis, we consider the sodic group as the reference or regular observations, to identify the remaining curves as outliers. The optimization parameters follow the same selection process as before. The subset size $h$ is set to $\lceil 0.5n \rceil$ and the regularization parameter $\alpha$ is chosen through automated selection, resulting in $\alpha = 2$.

In the right plot of Figure \ref{fig:epxma}, the resulting $\alpha$-Mahalanobis distances are displayed, with colors representing the grouping structure of the data. The vertical black line indicates the theoretical cutoff \ref{cor:cor_2}. Based on this analysis, the second, third, and fourth group are identified as outliers. Additionally, a few curves from the majority group are also detected as outlying. Further investigation revealed that some of the spectra within the sodic glass vessels were recorded under different experimental conditions. This discrepancy may account for the atypical curves observed within the sodic group.
  
\section{Summary and conclusions}

In this paper, we presented the minimum regularized covariance trace (MRCT) estimator, a novel method for robust covariance estimation and functional outlier detection. To handle outlying observations, we adopted a subset-based approach that favors the subsets that are the most central w.r.t the corresponding covariance, and where the centrality of the point is measured by the $\alpha$-Mahalanobis distance \eqref{def:alpha_mah}. The approach results with the fast-MCD \cite{rousseeuw1999} type algorithm, in which the notion of standard Mahalanobis distance is replaced by $\alpha$-Mahalanobis distance. Consequentially, the robust estimator of the $\alpha$-Mahalanobis distance is obtained, thus further allowing for robust outlier detection. The method automates outlier detection by providing theoretical cutoff values based on additional distributional assumptions. An additional advantage of our approach is its ability to handle data sets with a high number of observed time points ($p \gg n$) without requiring preprocessing or dimension reduction techniques such as smoothing.  This is a consequence of the fact that the $X_\mathrm{st}^\alpha$ is a smooth approximation (obtained via the Tikhonov regularization) of the solution $Y$ of ill-posed linear problem $C^{1/2}Y=X$,  where the amount of smoothening is determined by a (univariate) regularization parameter $\alpha>0$. Therefore, a certain amount of smoothening is in fact done within the procedure itself.  
Additionally, the typically challenging task of choosing the Tikhonov regularization parameter $\alpha$ is automated. We demonstrate the importance of selecting an appropriate value for $\alpha$, which determines the optimal subset, striking a balance between noise exclusion and preservation of signal components. Moreover, we provided the straightforward adaptation of the method to sparsely observed data, where the approach is to represent the data on a fixed basis and directly work with the finite matrix of basis coefficients, as opposed to the smoothed functions, further minimizing the approximation error. 
Overall, the MRCT estimator offers distinct advantages in terms of robust covariance estimation and automated outlier detection in functional data analysis, as demonstrated in the simulation study in Section \ref{sec:Simulations}.

As a part of future research, we will study the behavior and efficacy of different variants of Tikhonov regularization. While our paper focused on the "standard form" with a simple smoothness penalty of $\alpha ||Y||^2$, future investigations could encompass a broader scope by incorporating a derivative operator, denoted as $\alpha||LY||^2$, where $L$ represents a suitable operator; see e.g. \cite{golub1999}. Another avenue for exploration lies in extending the approach to accommodate also the case of multivariate functional data.

\section*{Acknowledgement}
The authors acknowledge support from the Austrian Science Fund (FWF), project number I 5799-N.

\section*{Disclosure statement}

The authors report there are no competing interests to declare.

\appendix
\section{Additional simulation results}\label{sec:appendix1}
\begin{figure}[h]
\centering
\subfloat[ $\alpha\in (0.001,0.1)\,$, $\lambda\in (10^{-3},0.1)$]{\includegraphics[width = 0.45\linewidth]{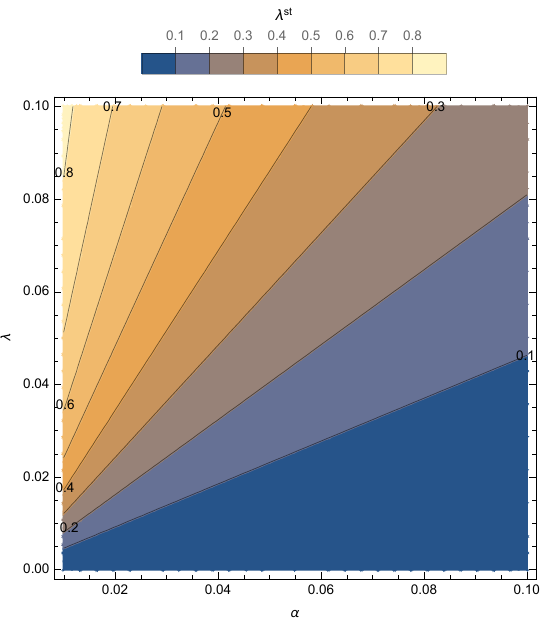}} 
\subfloat[ $\alpha\in (0.1,20)\,$, $\lambda\in (10^{-2},100)$]{\includegraphics[width = 0.45\linewidth]{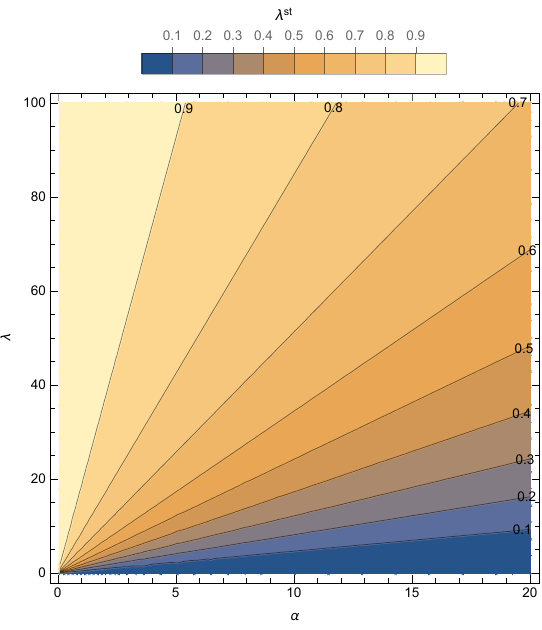}}
    \caption{Contour plot of eigenvalues $\lambda_{\mathrm{st}}^{\alpha}$ of $C(X_{\mathrm{st}}^{\alpha})$, for different combinations of smoothing parameter $\alpha>0$ and eigenvalues $\lambda>0$ of $C$.
    }
        \label{fig:fig1}
\end{figure}
 Figure~\ref{fig:fig1} shows the impact of $\alpha$ to eigenvalues of $C(X_{\mathrm{st}}^{\alpha})$, for a range of eigenvalues of $C$. Especially, it illustrates that for small $\alpha$, 
the value $\lambda_{i,\mathrm{st}}^{\alpha}\approx 1$, if $\lambda_i$ is reasonably large. On the other hand, if $\alpha$ is too large, in case of large and diverse $\lambda_i$, the corresponding values of $\lambda_i^\mathrm{st}$ might be too small and not "close enough to each other", thus diminishing some of the signal components.  Similarly, in the case of very small eigenvalues, too small values of $\alpha$ might make small, "irrelevant" eigenvalues large and diverse, thus magnifying noise.\\ 

In Section~\ref{sec:Simulations}, a simulation study is performed to evaluate the sensitivity of Algorithm \ref{alg:Functional} to the initial approximation. The simulation study utilized the regularization parameter determined by Algorithm~\ref{alg:alpha}.
 The results reveal that selecting the appropriate value of $\alpha$ can significantly reduce this sensitivity. 
 Consequently, it is observed that the choice of $H_{\mathrm{opt}}\in\mathcal{X}$ has minimal influence on the final outcomes, including the covariance estimate and outlier detection. To quantify this, the first row of Table \ref{tab:overlap} presents the average overlap between the obtained subsets using 100 different initial approximations and the chosen optimal subset $H_{\mathrm{opt}}$. The overlap is calculated as $\displaystyle O_1 := \frac{1}{h}|H\cap H_{\mathrm{opt}}|$, where $H$ represents the obtained output $H=H(H_0)\in\mathcal{X}$ of Algorithm~\ref{alg:Functional} for a randomly initiated $H_0$. The results demonstrate that the overlap is close to 1, indicating a high level of consistency between the obtained subsets and $H_{\mathrm{opt}}$. Additionally, the second row of 
 Table!\ref{tab:overlap} gives the overall overlap between all obtained subsets, where for 100 randomly initiated $H_0^i$, $H^i=H^i(H_0^i)\in\mathcal{X}$, $i=1,\dots,100$, the overall overlap is calculated as $\displaystyle O_2 := \frac{1}{h}|\cap_{i=1}^{100} H^i|$. 

\begin{table}[h] 
\centering
\begin{tabular}{llll} \hline
   & Model 1 & Model 2 & Model 3 \\  \hline 
$O_1$ & 0.981   & 0.992   & 0.992   \\
$O_2$ & 0.954   & 0.980   & 0.979 \\ \hline
\end{tabular}
\vspace{2ex}
\caption{Overlap ratio between optimal subset $H_{\text{opt}}$ and all obtained subsets $(O_1)$ as well as the ratio of intersection between all subsets $(O_2)$. Mean values where calculated over $100$ iterations on $100$ random initializations.}
\label{tab:overlap}
\end{table}

\section{Proofs}\label{sec:appendix2}
\begin{proof}[Proof of Lemma~\ref{lem:fixed_point}]
Let us first define the set  $\mathcal{U}_h=\{(w_1,\dots,w_n)\in[0,1]^n:\,\sum_{i=1}^nw_i=h\}$ and the function $g:\mathcal{U}_h\times\mathcal{U}_h\to \mathcal{U}_h$ with 
$$
g(\textbf{w};\textbf{w}_0)=\frac{1}{\sum_{i=1}^n w_i}\sum_{i=1}^nw_i^2\|\hat{C}_{\textbf{w}_0}^{1/2}(\hat{C}_{\textbf{w}_0}+\alpha I)^{-1} (X_i-\bar{X}_{\textbf{w}_0}) \|^2,
$$
where for $Y\in L^2(I)$, $\hat{C}_{\textbf{w}_0}Y=\frac{1}{\sum_{i=1}^n{w_{0,i}}}\sum_{1=1}^n w_{0,i}\langle(X_i-\bar{X}_{\textbf{w}_0}),Y\rangle (X_i-\bar{X}_{\textbf{w}_0})$, and $\bar{X}_{\textbf{w}_0}=\frac{1}{\sum_{i=1}^n{w_{0,i}}}\sum_{i=1}^nw_iX_i$. Let us first show that $\mathcal{U}_h$ is a convex set. Let $\textbf{w}_1,\,\textbf{w}_2\in\mathcal{U}_h$ and $\lambda\in(0,1)$, and define $\textbf{w}=\lambda\textbf{w}_1+(1-\lambda)\textbf{w}_2$. Then, $w_i=\lambda w_{1,i}+(1-\lambda) w_{2,i}\in [0,1]$, and $\sum_{i=1}^hw_i=\lambda h+(1-\lambda)h=h$, implying $\textbf{w}\in\mathcal{U}_h$.

Define now $d_i^2=\|\hat{C}_{\textbf{w}_0}^{1/2}(\hat{C}_{\textbf{w}_0}+\alpha I)^{-1} (X_i-\bar{X}_{\textbf{w}_0}) \|^2$, and order the obtained quantities as $d_{i_1}\leq\cdots\leq d_{i_n}$. Then observe that if we were to optimize $g$ w.r.t. $\textbf{w}$, keeping $\textbf{w}_0$ as a fixed parameter, the result of the given optimization problem is given explicitly with $w_i=1$, if $i\in\{i_1,\dots,i_h\}$, and $0$ otherwise. Such $\textbf{w}$ uniquely defines a corresponding set of observations with $H=\{i_1,\dots,i_h\}$. Define further 
$f:\mathcal{U}_{h}\to \mathcal{U}_{h}$ with 
$f_\textbf{w}(\textbf{w}_0)=\argmin_{\textbf{w}\in \mathcal{U}_{h}}g(\textbf{w};\textbf{w}_0)$. Then any fixed point $\textbf{w}_0$ of $f_\textbf{w}$, due to the same reasoning as above,  uniquely defines a corresponding set of observations $H_0$. Hence, if we show that $f_\textbf{w}$ has a fixed point, then that fixed point uniquely defines a fixed point of $f$. Now, since $g$ is a continuous function (as a composition of continuous functions) on its domain, Berge's maximum theorem~\citep{Berge1963} implies that $f_\textbf{w}$ is continuous as well. Finally, as $f_\textbf{w}$ is a continuous mapping from a compact and convex set $\textbf{U}_h$ into itself (not necessarily surjective), Brouwer's fixed-point theorem  \citep{Brouwer1911} then implies that $f_\textbf{w}$ has at least one fixed point $\textbf{w}_0\in\mathcal{U}_h$ that, for the reasons stated above, defines a fixed point $H_0$ of $f$. 
\end{proof}

\begin{proof}[Proof of Proposition~\ref{prop:prop_1}]
 \begin{align*}
     |\|X_{\mathrm{st},H_0}^{\alpha,k_{H_0}}\|-\|X_{\mathrm{st}}^\alpha\||&\leq \|\sqrt{k_{H_0}}\hat{C}_{H_0}^{1/2}({k_{H_0}}\hat{C}_{H_0}+\alpha I)^{-1}(X-\bar{X}_{H_0})-{C}^{1/2}({C}+\alpha I)^{-1}X\|\\
     &\leq \|\sqrt{k_{H_0}}\hat{C}_{H_0}^{1/2}({k_{H_0}}\hat{C}_{H_0}+\alpha I)^{-1}-{C}^{1/2}({C}+\alpha I)^{-1}\|_{op}\|X\|\\
     &+\|\sqrt{k_{H_0}}\hat{C}_{H_0}^{1/2}\|_{op}\|({k_{H_0}}\hat{C}_{H_0}+\alpha I)^{-1}\|_{op}\|\bar{X}_{H_0}\|,
 \end{align*} 
 where $\|\cdot\|_{op}$ denotes the operator norm.
First, observe that 
\begin{align*}
     &\|\sqrt{k_{H_0}}\hat{C}_{H_0}^{1/2}({k_{H_0}}\hat{C}_{H_0}+\alpha I)^{-1}-{C}^{1/2}({C}+\alpha I)^{-1}\|_{op}\leq \\
      &\|\sqrt{k_{H_0}}\hat{C}_{H_0}^{1/2}({k_{H_0}}\hat{C}_{H_0}+\alpha I)^{-1}-\sqrt{k_{H_0}}\hat{C}_{H_0}^{1/2}(C+\alpha I)^{-1}\|_{op}
      +\|\sqrt{k_{H_0}}\hat{C}_{H_0}^{1/2}(C+\alpha I)^{-1}-{C}^{1/2}({C}+\alpha I)^{-1}\|_{op}\leq\\
      & \|\sqrt{k_{H_0}}\hat{C}_{H_0}^{1/2}\|_{op}\|({k_{H_0}}\hat{C}_{H_0}+\alpha I)^{-1}-(C+\alpha I)^{-1}\|_{op}+\|\sqrt{k_{H_0}}\hat{C}_{H_0}^{1/2}-C^{1/2}\|_{op}\|(C+\alpha I)^{-1}\|_{op}.
\end{align*}
 Since $k_{H_0}\hat{C}_{H_0}\to_{a.s.}C$, there exists a bound on the operator norm of $k_{H_0}\hat{C}_{H_0}$. 
 As $\|k_{H_0}\hat{C}_{H_0}-C \|_{op}=\|(k_{H_0}\hat{C}_{H_0}+\alpha I)-(C+\alpha I)\|_{op}\to_{a.s.} 0$, and $\sup_n\|(k_{H_0}\hat{C}_{H_0}+\alpha I)^{-1}\|_{op}\leq \alpha^{-1}$;
 ,  Lemma 8 from \cite{berrendero2020} implies that $\|(k_{H_0}\hat{C}_{H_0}+\alpha I)^{-1}-(C+\alpha I)^{-1}\|_{op}\to_{a.s.} 0$. Thus, $\|\sqrt{k_{H_0}}\hat{C}_{H_0}^{1/2}\|_{op}\|({k_{H_0}}\hat{C}_{H_0}+\alpha I)^{-1}-(C+\alpha I)^{-1}\|_{op}\to_{a.s.} 0$. 
 Furthermore, as $k_{H_0}\hat{C}_{H_0}\to_{a.s.}C$, $k_{H_0}\hat{C}_{H_0}$ is bounded and the square root is a continuous function, $\|\sqrt{k_{H_0}}\hat{C}_{H_0}^{1/2}-C^{1/2}\|_{op}\to_{a.s.} 0$. Finally, since for every $Y\in L^2(I)$,  $(C+\alpha_I)^{-1}Y\leq \alpha^{-1}$, $\|\sqrt{k_{H_0}}\hat{C}_{H_0}^{1/2}-C^{1/2}\|_{op}\|(C+\alpha I)^{-1}\|_{op}\to 0$.
 Due to  the assumption of the statement, $\|\bar{X}_{H_0}\|\to_{a.s} 0$, finally implying that $\|\sqrt{k_{H_0}}\hat{C}_{H_0}^{1/2}\|_{op}\|({k_{H_0}}\hat{C}_{H_0}+\alpha I)^{-1}\|_{op}\|\bar{X}_{H_0}\|\to_{a.s} 0$, thus proving the claim.
\end{proof}

\begin{proof}[Proof of Corollary~\ref{cor:cor_2}]
   Proposition~\ref{prop:prop_2} implies that  $\|X_{\mathrm{st},H_0}^{\alpha,k_{H_0}}\|\to_{a.s} \mathrm{MD}_\alpha(X,0)$, as $n\to\infty$. Proposition~6 in~\cite{berrendero2020} states that under the assumption of the corollary, $\mathrm{MD}_\alpha(X,0)\sim  \sum_{i=1}^\infty\frac{\lambda_i^2}{(\lambda_i+\alpha)^2}Y_i$, where $Y_i$, $i=1,2\dots$ are independent random variables from $\chi^2(1)$, and  $\lambda_i$, $i=1,2\dots$ are eigenvalues of $C$.  
\end{proof}

\begin{proof}[Proof of Proposition~\ref{prop:prop_3}]
Following the proof of Proposition~\ref{prop:prop_1}
 \begin{align*}
     |\|X_{\mathrm{st},H_0}^{\alpha_n,k_{H_0}}\|-\|X_{\mathrm{st}}^\alpha\||&\leq \|\sqrt{k_{H_0}}\hat{C}_{H_0}^{1/2}({k_{H_0}}\hat{C}_{H_0}+\alpha_n I)^{-1}(X-\bar{X}_{H_0})-{C}^{1/2}({C}+\alpha I)^{-1}X\|\\
     &\leq \|\sqrt{k_{H_0}}\hat{C}_{H_0}^{1/2}({k_{H_0}}\hat{C}_{H_0}+\alpha_n I)^{-1}-{C}^{1/2}({C}+\alpha I)^{-1}\|_{op}\|X\|\\
     &+\|\sqrt{k_{H_0}}\hat{C}_{H_0}^{1/2}\|_{op}\|({k_{H_0}}\hat{C}_{H_0}+\alpha_n I)^{-1}\|_{op}\|\bar{X}_{H_0}\|,
 \end{align*} 
where
\begin{align*}
     &\|\sqrt{k_{H_0}}\hat{C}_{H_0}^{1/2}({k_{H_0}}\hat{C}_{H_0}+\alpha I)^{-1}-{C}^{1/2}({C}+\alpha I)^{-1}\|_{op}\leq \\
      & \|\sqrt{k_{H_0}}\hat{C}_{H_0}^{1/2}\|_{op}\|({k_{H_0}}\hat{C}_{H_0}+\alpha I)^{-1}-(C+\alpha I)^{-1}\|_{op}+\|\sqrt{k_{H_0}}\hat{C}_{H_0}^{1/2}-C^{1/2}\|_{op}\|(C+\alpha I)^{-1}\|_{op} 
\end{align*}
and $\|\cdot\|_{op}$ denotes the operator norm. As discussed, since $k_{H_0}\hat{C}_{H_0}\to_{a.s.}C$, there exists a bound on the operator norm of $k_{H_0}\hat{C}_{H_0}$. Furthermore,  $\|k_{H_0}\hat{C}_{H_0}-C \|\leq\|(k_{H_0}\hat{C}_{H_0}+\alpha_n I)-(C+\alpha I)\|+|\alpha_n-\alpha|\to_{a.s.} 0$, and since the operators involved are bounded and invertible,  $\|(k_{H_0}\hat{C}_{H_0}+\alpha_n I)^{-1}-(C+\alpha I)^{-1}\|\to_{a.s.} 0$;  see (Lemma 8,  \cite{berrendero2020}). The rest of the proof follows the one of Proposition~\ref{prop:prop_1}.
\end{proof}

\begin{proof}[Proof of Proposition~\ref{prop:Mah_distance_in_a_basis}]
For simplicity of the notation, we will drop the subscript $H_0$ from the eigenvalues and eigen-functions of $\hat{C}_{H_0}$.  We start by jointly proving statements (i) and (ii). First observe that since by assumption $X_i\in\mathrm{span}(\phi_1,\dots,\phi_M)$, $i=1,\dots,m$, then also $\bar{X}_{H_0}\in\mathrm{span}(\phi_1,\dots,\phi_M)$. Furthermore, since for any $X\in L^2(I)$ the sample covariance operator satisfies $\hat{C}_{H_0}=\frac{1}{h}\sum_{i\in H_0}\langle X_i-\bar{X}_{H_0},X\rangle (X_i-\bar{X}_{H_0})$, $\hat{C}_{H_0}:L_2(I)\to  \mathrm{span}(\phi_1,\dots,\phi_M)$. Finally, since $\hat{C}_{H_0}\hat{\psi}_i=\hat{\lambda}_i\hat{\psi}_i$, then  $\hat{\psi}_i\in \mathrm{span}(\phi_1,\dots,\phi_M)$ thus implying that $\hat{\psi}_i=\textbf{u}_i' \boldsymbol{\Phi}$, for some vector $\textbf{u}_i$, $i=1,\dots,M$. Furthermore, since $\boldsymbol{\Phi}$ is orthonormal, then so are $\textbf{u}_1,\dots,\textbf{u}_M$.  The following is now equivalent for $j=1,\dots,M$;
\begin{align*}
\hat{C}_{H_0}\hat{\psi}_j=\hat{\lambda}_j\hat{\psi}_j &\iff \frac{1}{h}\sum_{i\in H_0}\langle X_i-\bar{X}_{H_0},\hat{\psi}_j\rangle ( X_i-\bar{X}_{H_0})=\hat{\lambda}_j\hat{\psi}_j\\
&\iff \frac{1}{h}\sum_{i\in H_0}\langle (\textbf{e}_i-\frac{1}{h}\mathbf{1}_{H_0})'\textbf{C}\boldsymbol{\Phi}, \boldsymbol{\Phi}'\textbf{u}_j\rangle(\textbf{e}_i-\frac{1}{h}\mathbf{1}_{H_0})'\textbf{C}\boldsymbol{\Phi}=\hat{\lambda}_j\textbf{u}_j'\boldsymbol{\Phi}\\
&\iff \boldsymbol{\Phi}'\textbf{C}'\frac{1}{h}\sum_{i\in H_0}(\textbf{e}_i-\frac{1}{h}\mathbf{1}_{H_0})(\textbf{e}_i-\frac{1}{h}\mathbf{1}_{H_0})'\textbf{C}\textbf{u}_j=\hat{\lambda}_j\boldsymbol{\Phi}'\textbf{u}_j\\
&\iff \boldsymbol{\Phi}'\textbf{C}'\frac{1}{h}(\textbf{I}_{n,H_0}-\frac{1}{h}\textbf{J}_{n,H_0})'\textbf{C}\textbf{u}_j=\hat{\lambda}_j\boldsymbol{\Phi}'\textbf{u}_j\\
&\iff \textbf{C}'\frac{1}{h}(\textbf{I}_{n,H_0}-\frac{1}{h}\textbf{J}_{n,H_0})'\textbf{C}\textbf{u}_j=\hat{\lambda}_j\textbf{u}_j,
\end{align*}
where the last equivalence holds since $\boldsymbol{\Phi}(t)'\textbf{C}'\frac{1}{h}(\textbf{I}_{n,H_0}-\frac{1}{h}\textbf{J}_{n,H_0})'\textbf{C}\textbf{u}_j=\hat{\lambda}_j\boldsymbol{\Phi}(t)'\textbf{u}_j$, for every $t\in I$, thus proving statements (i) and (ii). We now use the obtained results to prove statement (iii). 
\begin{align*}
    d_{i,H_0,\alpha}&=\|\hat{C}_{H_0}^{1/2}(\hat{C}_{H_0}+\alpha I)^{-1}(X_i-\bar{X}_{H_0})\|^2\\
    &=\sum_{j=1}^M\frac{\hat{\lambda}_j}{(\hat{\lambda}_j+\alpha)^2}\langle\hat{\psi}_j,X_i-\bar{X}_{H_0}\rangle^2\\
    &=\sum_{j=1}^M\frac{\hat{\lambda}_j}{(\hat{\lambda}_j+\alpha)^2}\langle\textbf{u}_j'\boldsymbol{\Phi},\boldsymbol{\Phi}'\textbf{C}'(\textbf{e}_i-\frac{1}{h}\mathbf{1}_{H_0})\rangle^2\\
    &=\sum_{j=1}^M\frac{\hat{\lambda}_j}{(\hat{\lambda}_j+\alpha)^2}\left((\textbf{e}_i-\frac{1}{h}\mathbf{1}_{H_0})'\textbf{C}\textbf{u}_j \right)^2\\
    &=(\textbf{e}_i-\frac{1}{h}\mathbf{1}_{H_0})'\textbf{C}\sum_{j=1}^M\frac{\hat{\lambda}_j}{(\hat{\lambda}_j+\alpha)^2}\textbf{u}_j\textbf{u}_j'\textbf{C}'(\textbf{e}_i-\frac{1}{h}\mathbf{1}_{H_0})\\
    &=(\textbf{e}_i-\frac{1}{h}\mathbf{1}_{H_0})'\textbf{C}\tilde{\textbf{C}}_{H_0}(\tilde{\textbf{C}}_{H_0}+\alpha\textbf{I}_M)^{-2}\textbf{C}'(\textbf{e}_i-\frac{1}{h}\mathbf{1}_{H_0}),
\end{align*}
where the final equality holds since $(\hat{\lambda}_j,\textbf{u}_j)$ is an eigen-pair of $\tilde{\textbf{C}}_{H_0}$. Observe first that  $(\textbf{I}_{n,H_0}-\frac{1}{h}\textbf{J}_{n,H_0})$ is a symmetric idempotent matrix, further making $\tilde{\textbf{C}}_{H_0}$ a symmetric, positive semi-definite matrix. Thus, $(\tilde{\textbf{C}}_{H_0}+\alpha \textbf{I}_n)$ is a positive definite symmetric matrix, whose inverse is then well defined. Therefore, $(\frac{\hat{\lambda}_j}{\hat{\lambda}_j+\alpha},\textbf{u}_j)$ is an eigen-pair of $\tilde{\textbf{C}}_{H_0}(\tilde{\textbf{C}}_{H_0}+\alpha\textbf{I}_M)^{-2}$, $j=1,\dots, M$.
\end{proof}

\bibliographystyle{unsrtnat}
\bibliography{references}  

\end{document}